\documentclass[twocolumn]{aastex631}
\usepackage{amsmath,amssymb,latexsym}
\newcommand{\msunyr}{M_{\odot}~{\rm yr^{-1}}}

\newcommand{\msun}{{\mbox M}_\odot}
\newcommand{\rsun}{{\mbox R}_\odot}
\shorttitle{Kirihara et~al.}
\shortauthors{Kirihara et~al.}

\begin{document}

\title{Merger Conditions of Population~III Protostar Binaries}

\author{Takanobu Kirihara}
\affiliation{Center for Computational Sciences, University of Tsukuba, Tennodai 1-1-1, Tsukuba, Ibaraki 305-8577, Japan}
\email{kirihara@ccs.tsukuba.ac.jp}

\author{Hajime Susa}
\affiliation{Department of Physics, Konan University, Okamoto, Kobe, Hyogo 658-8501, Japan}


\author{Takashi Hosokawa}
\affiliation{Department of Physics, Graduate School of Science, Kyoto University, Sakyo-ku, Kyoto 606-8502, Japan}

\author{Tomoya Kinugawa}
\affiliation{Shinshu University, 4-17-1 Wakasato, Nagano 380-8553, Japan}
\affiliation{Research Center for the Early Universe, Graduate School of Science, University of Tokyo, 7-3-1 Hongo, Bunkyo-ku, Tokyo 113-0033, Japan}



\begin{abstract}


Massive close binary stars with extremely small separations have been observed, and they are possible progenitors of gravitational-wave sources. 
The evolution of massive binaries in the protostellar accretion stage is key to understanding their formation process. 
We, therefore, investigate how close the protostars, consisting of a high-density core and a vast low-density envelope, can approach each other but not coalesce. 
To investigate the coalescence conditions, we conduct smoothed particle hydrodynamics simulations following the evolution of equal-mass binaries with different initial separations. 
Since Population (Pop) I and III protostars have similar interior structures, we adopt a specific Pop~III model with the mass and radius of $7.75\;M_{\odot}$ and $61.1\;R_{\odot}$ obtained by the stellar evolution calculations. 
Our results show that the binary separation decreases due to the transport of the orbital angular momentum to spin angular momentum. 
If the initial separation is less than about 80 per~cent of the sum of the protostellar radius, the binary coalesces in a time shorter than the tidal lock timescale. 
The mass loss up to the merging is $\lesssim 3$ per~cent. 
After coalescence, the star rotates rapidly, and its interior structure is independent of the initial separation. 
We conclude that there must be some orbital shrinking mechanism after the protostars contract to enter the zero-age main-sequence stage. 

\end{abstract}



\section{Introduction}
\label{sec:intro}

Population III (hereafter Pop~III) stars are believed to form in primordial gas clouds hosted by dark matter minihalos with $\sim 10^{5-6}\;M_{\odot}$ \citep[e.g.,][]{Haiman1996, Tegmark1997, Nishi1999, Abel2002, Yoshida2003}. 
Their formation process after the cloud collapse has been investigated in previous theoretical works. Recent numerical simulations have demonstrated that the circumstellar disks form and are gravitationally unstable to fragment, resulting in multiple protostars \citep[e.g.,][]{Greif2011,Wollenberg2020, Sugimura2020}. 
These studies show that many fragments form in a circumstellar disk, and a large fraction of the protostars have experienced a binary phase \citep{Susa2019,Shima21,Chiaki2022}.

It is highly uncertain that close protostellar binaries survive until becoming main-sequence binary stars. 
In order to reduce the computational cost, a sink particle approach has been used to simulate the evolution of the protostars and the circumstellar disks \citep{Stacy2010, Stacy2012, Clark2011, Smith2011, Susa2013, Susa2014,Sugimura2020,Sharda2020}. 
For example, in smoothed particle hydrodynamics (SPH) simulations, one often inserts sink particles representing accreting protostars when the local density exceeds a threshold value. 
\citet{Greif2011} and \citet{Prole2022} pointed out that the number of fragments depends on the sink radius and sink-creation criteria. 
Such studies also suffer from uncertain merger conditions between sink particles, which should differ from actual conditions between accreting protostars. 
Stellar evolution calculations predict that an accreting Pop~III protostar swells and consists of a high-density core and a vast low-density envelope \citep[e.g.,][]{Stahler86,Omukai2003}. 
Therefore, it is necessary to investigate how close the protostars can approach each other but not coalesce, considering their expected interior structures. 
Clarifying the coalescence conditions of protostars should reveal the physical coalescence conditions in numerical simulations using sinks.

Regarding massive close binaries in the present-day universe, our knowledge on their formation process is also limited in spite of informative observations. 
\citet{Sana2012} reported the binarity and multiplicity of O-type main-sequence stars from the observation of six galactic young open clusters. 
They identified that over half of all O-type stars are binaries. 
The sample contains several close binaries with orbital periods of less than 10~days, which corresponds to the binary separation of several tens to one hundred $\rsun$. 
The recently discovered VFTS~352 is a remarkably tight binary with stellar masses of 28.63~$\msun$ and 28.85~$\msun$, and an orbital semimajor axis is 17.55~$\rsun$. 
This system is expected to be in the contact phase \citep{Almeida2015}.
In contrast, theoretical studies also predict that Pop I massive protostars should have large radii ($\sim 100~\rsun$) under rapid accretion normally supposed \citep[e.g.][]{McKee03,Yorke2008,Hosokawa2009,Hosokawa2010,Haemmerle16}. 
It is still under debate how such close binaries form \citep[e.g.][]{Krumholz2007,Moe2018,Tokovinin2020,Harada21}.

The advent of the gravitational-wave astronomy draws great attention to the massive close binary formation. 
Formation of Pop~III binaries is of particular interest since they are potential candidates for gravitational-wave source progenitors. 
If Pop~III stars have stellar masses $>50 M_{\odot}$, they evolve to red giants, and such binaries easily coalesce by extracting angular momentum in the common envelope phase, even if they are not close binaries. 
On the other hand, if the stellar mass is $<50 M_{\odot}$, they evolve to blue giants, and the stellar radius does not enlarge so much, indicating that close binaries are more likely to coalesce \citep[e.g.][]{Kinugawa2014,Inayoshi2017,Tanikawa2021}. 
If they are born as sufficiently close main-sequence binaries, they could evolve to black hole binaries that will merge within the Hubble time. 
Previous studies often assume that Pop~III main-sequence close binaries form to investigate further binary evolution leading to the black hole binary mergers \citep{Kinugawa2014,Kinugawa2016,Kinugawa2020,Hartwig2016,Belczynski2017,Sugimura2020,Liu2020a,Tanikawa2021,Tanikawa2021a,Tanikawa2022,Costa2023,Santoliquido2023}. 
However, it is still highly uncertain whether such tight binaries appear after the protostellar accretion stage. 
Therefore, it is helpful to investigate how close the protostars can approach each other without coalescing.

Numerical simulations of main-sequence stellar mergers have been carried out by several authors so far \citep{Seidl1972,Benz1987,Benz1992,Davies1994,Freitag2005,Suzuki2007}. 
One major difference between main sequence stars and Pop~III protostars is that the latter consist of a high-density core and a vast low-density envelope \citep[e.g.,][]{Omukai2003}. 
Besides, dynamical simulations of the coalescence of protostar binaries have not been performed to date. 
Since the protostars of Pop~I and Pop~II stars have a similarly large radius as the Pop~III protostars \citep[e.g.][]{Hosokawa2009}, the present simulation is beneficial not only for the primordial environment but also for the present-day environment. 

In this study, we investigate the merger process of two protostars by means of SPH simulations. 
This paper is organized as follows. 
We describe the details of our SPH simulations in Section \ref{sec:method}. 
We also explain the initial conditions and simulation setup. 
We show the results of the numerical simulations in Section \ref{sec:result}. 
We discuss the timescale of coalescence in Section \ref{sec:discussion} and provide concluding remarks in Section \ref{sec:summary}.

\section{Numerical Method and Initial Conditions}\label{sec:method}

In this work, we study the evolution and the merger condition of a Pop~III protostellar binary. 
We carry out SPH simulations to calculate the binary evolution of the protostars. The details of the code implementations are described in \citet{Susa2004} and \citet{Susa2006}. 

\begin{figure}
    \centering
    \includegraphics[width=\linewidth]{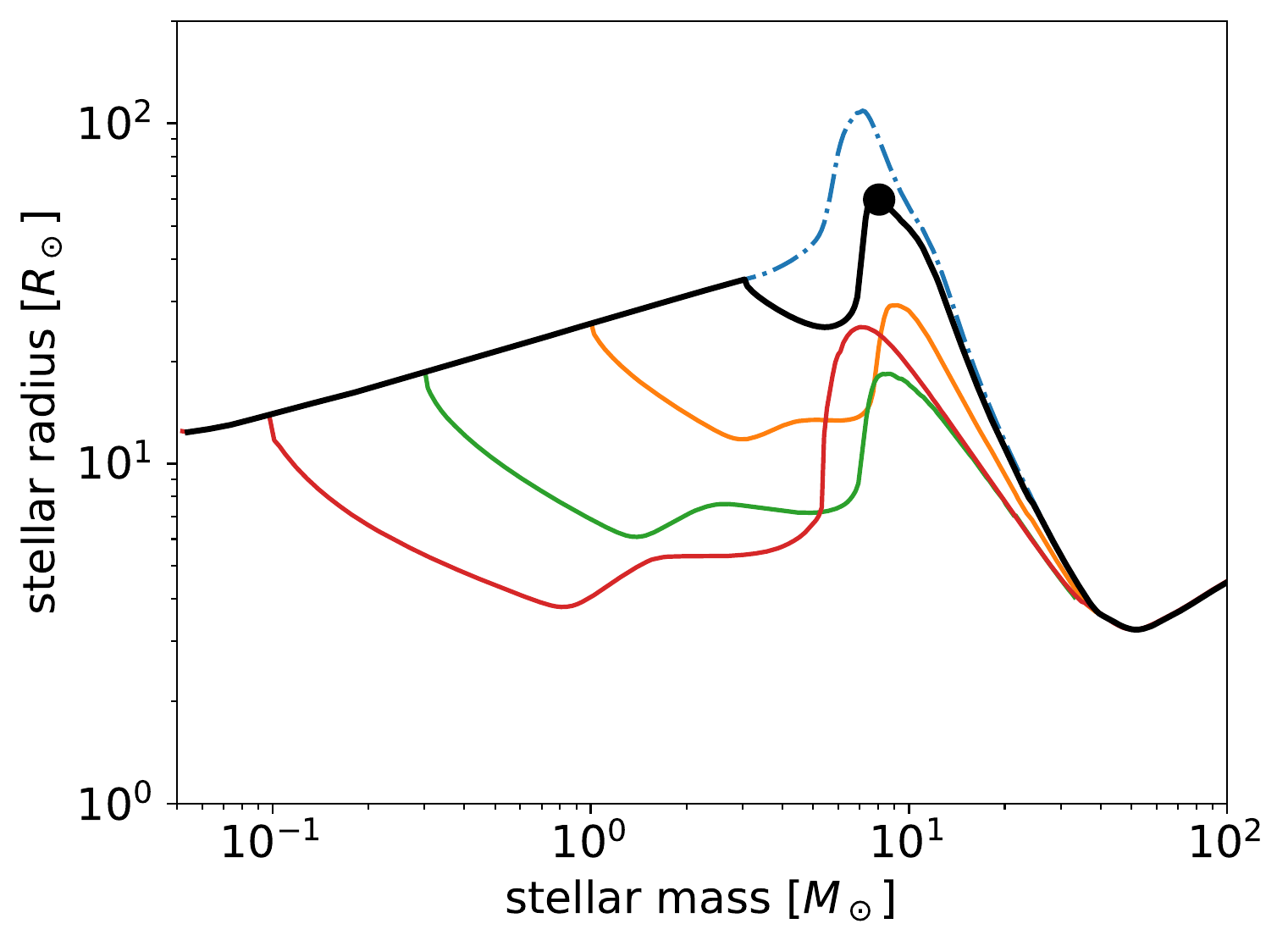}
    \caption{
    Results of the Pop III protostellar evolution calculations. 
    Plotted are variations of the stellar radius as a protostar accretes the gas at the rate of $10^{-3}\;\msunyr$ with different geometries. 
    The blue dot-dashed line represents the case where the spherical geometry is assumed until the star accretes 100~$M_\odot$ of the gas. 
    The red, green, orange, and black solid lines represent the cases where the geometry changes to the disk accretion at $M_*=0.1\;M_{\odot}, 0.3\;M_{\odot}, 1\;M_{\odot},$ and 3$\;M_{\odot}$, respectively. 
    The filled circle marks the stellar model used for our SPH simulations. 
} 
    \label{fig:Hosokawa}
\end{figure}

To investigate the evolution of Pop~III protostar binaries, we conduct stellar evolution calculations solving the interior structure of accreting protostars, following \citet{Hosokawa2009} and \citet{Hosokawa2010}. We consider a specific case where a protostar accretes the gas at the constant rate of $10^{-3}\;\msunyr$, the typical value in the Pop~III star formation. 
As in \citet{Hosokawa2010}, we also consider different accretion geometries, the spherical and disk accretion, by changing the outer boundary condition for the protostellar models \citep[see also][]{Palla1992}. 
Figure~\ref{fig:Hosokawa} shows the resulting evolution of the protostellar radius. 
The red, green, orange, and black solid lines show the evolution when the accretion geometry changes from spherical accretion to disk accretion at $M_*=0.1\;\msun$, $0.3\;\msun$, $1\;\msun$, and 3$\;\msun$, respectively. 
We see that for all the cases the protostar tends to swell as the stellar mass approaches $\sim 10~\msun$. 
After this, the protostar eventually begins Kelvin–Helmholtz (KH) contraction and reaches the zero-age main sequence at $\simeq 50~\msun$. 
If the accretion geometry is switched earlier, the stellar radius during the swelling stage tends to be smaller. 
We adopt the stellar model at the epoch just after the swelling stage as the initial condition for our SPH simulations. 
This is the case of our maximal interest, as we expect that possible mergers between such swollen protostars should limit the formation of massive tight binary systems. 
We use a specific protostar model whose radius and mass are $R_*=61.1~\rsun$ and $M_*=7.75~\msun$, which corresponds to the model marked by the filled circle in Figure~\ref{fig:Hosokawa}. 
We discuss the possible effects of varying stellar models in Section~\ref{sec:discussion}.

\begin{figure}
    \centering
        \includegraphics[width=\linewidth]{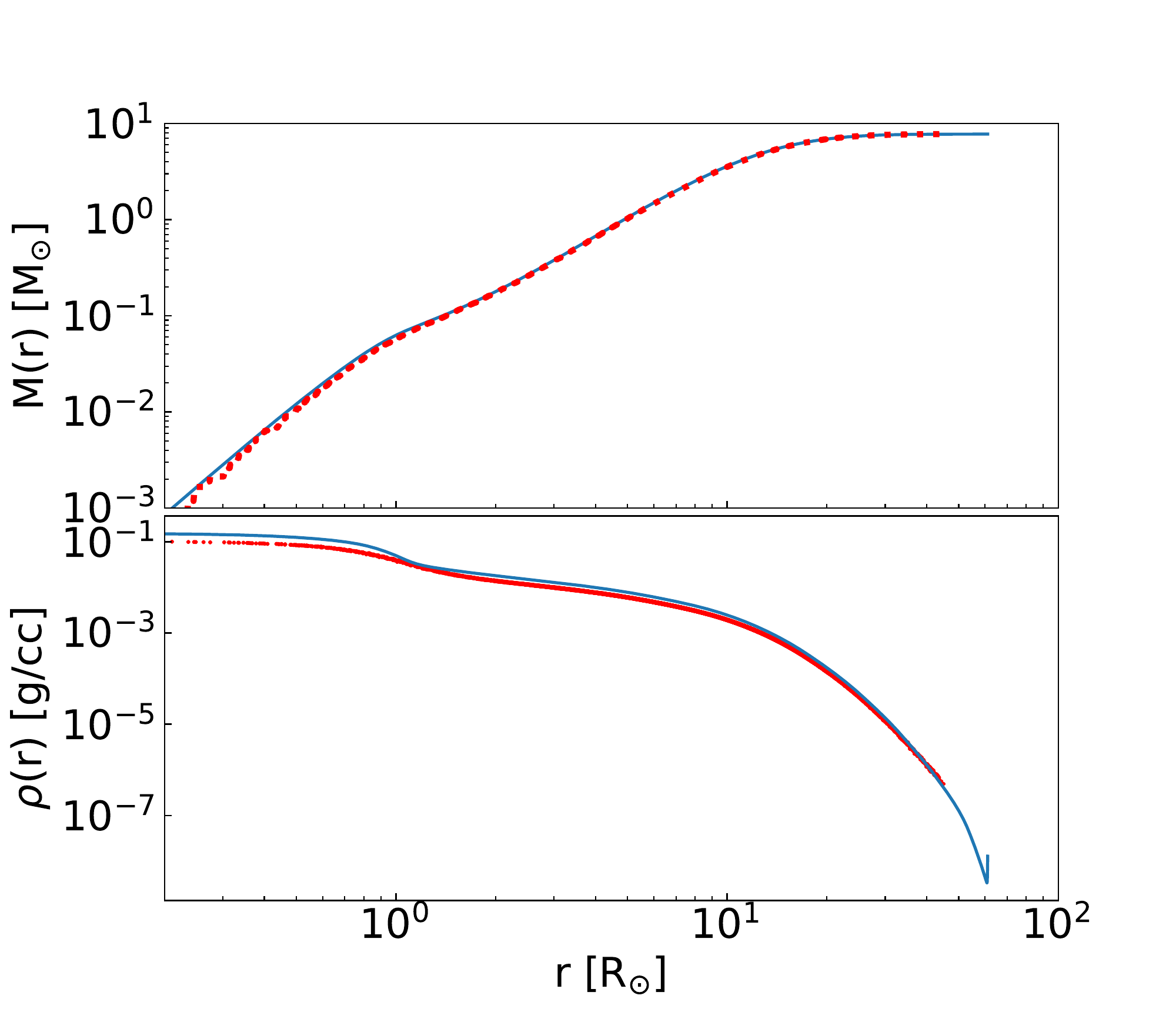}
    \caption{Mass distribution (upper panel) and density profile (lower panel) of the protostar model at the initial condition. 
    The red dotted line represents the 3D particle distribution after 5.2 $T_{\rm dyn}$ of relaxation. 
    The solid blue line shows the result of the stellar evolution calculation. 
}
    \label{fig:IC}
\end{figure}

In Figure~\ref{fig:IC}, we present the interior structure of the stellar model we consider. 
As indicated by the lower panel showing the density distribution, the stellar model has a two-layer structure, a high-density core ($r\lesssim R_{\odot}$) and a low-density envelope ($r\gtrsim R_{\odot}$). 
In this model, radiation pressure partially (a few tens per~cent in the envelope) contributes to the internal pressure of the protostar. 
In our calculations, instead of solving for the radiation process, we adjust the gas temperatures at each radius to be consistent with the pressure distribution obtained from the protostellar evolution calculations. 
We fix the average molecular weights of hydrogen and helium at fully ionized conditions. 
The gas consists of 75 per~cent hydrogen and 25 per~cent helium by mass, and the mean molecular weight is $\mu=16/27$.

We construct a 3D protostellar model based on the interior structure obtained by the stellar evolution calculation. 
We adopt the number of 32768 SPH particles to construct the protostellar model. 
The numerical viscosity of the SPH method allows us to introduce the dissipative terms needed for angular momentum transport.  
We return to this point in Section \ref{sec:timescale}. 
Owing to a limited spatial resolution with the particle approximation, the 3D model does not perfectly satisfy the hydrostatic equilibrium to cause some radial oscillations. 
We damp the spurious oscillations by conducting an SPH simulation for a specific time to relax the structure. 
Following previous studies \citep{Sato2015, Tanikawa2015}, we deliberately enhance the numerical viscosity to this end. 
We temporarily adopt $\alpha=20$ and $\beta=30$ (otherwise, $\alpha=1$ and $\beta=2$), where $\alpha$ and $\beta$ are control parameters of the viscosity in the standard formulation of the SPH method \citep[e.g.,][]{Monaghan1988,Hernquist1989}. 
We employ the model with $18$~days $(=5.2\;T_{\rm dyn})$ relaxation as the 3D protostellar model for the initial conditions, where $T_{\rm dyn}$ is the dynamical time of the original protostar.

Figure~\ref{fig:IC} shows that after the relaxation the 3D stellar model reproduces the overall mass and density distributions obtained by the 1D stellar evolution calculations. 
In the central region, the density is slightly lower due to the effect of the limited number of particles. 
The outer radius becomes 45~$R_{\odot}$ because the SPH particles cannot reside in a very rarefied region, where only 0.11 per~cent of the mass lies in the original protostellar model. 
The density of the outer region is so low that it has little effect on the coalescence process. 
We come back to this point in Section~\ref{subsec:spinrot}.

\begin{figure}
    \centering
    \includegraphics[width=\linewidth]{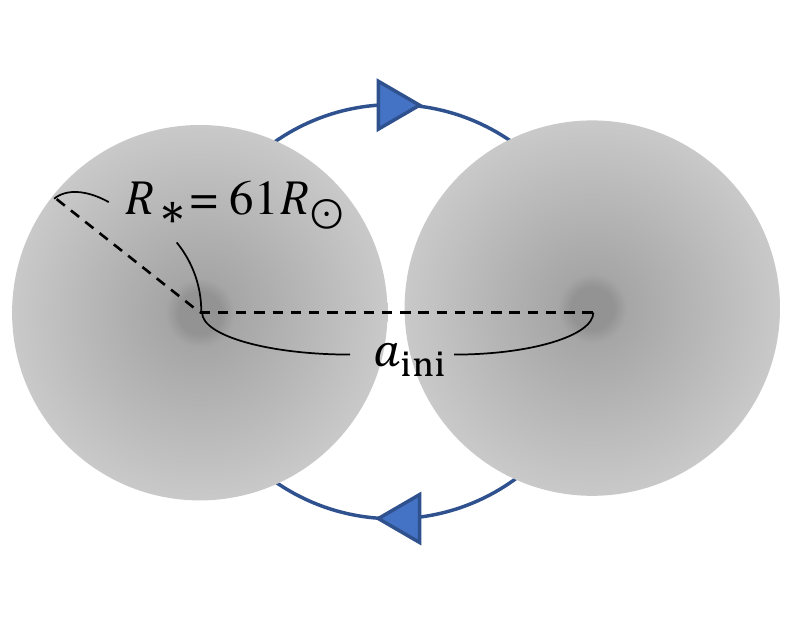}
    \caption{Schematic picture of the initial condition of our binary evolution simulations. 
    The gray spheres represent the protostars. 
    The blue circle with arrows indicates the binary protostars' orbit and direction of motion. 
} 
    \label{fig:schem1}
\end{figure}

As the initial condition of the SPH simulations, we place the two protostars separated by the distance $a_{\rm ini}$. 
Figure~\ref{fig:schem1} illustrates a schematic picture of the initial conditions of our simulations. 
The parameter $a_{\rm ini}$ takes values from 74~$R_{\odot}$ to 118~$R_{\odot}$ at equal intervals of  3~$R_{\odot}$ for 17 runs in total. 
We set the initial velocity of the two protostars assuming the circular orbit with their separation of $a_{\rm ini}$.

We neglect gas accretion onto the binaries because the accreting mass during the simulation duration is small enough. 
The rapid accretion can change the binary separation during long-time orbital motion. \citet{Sugimura2020} show in their long-term protostellar evolution simulation that the binary expands due to the gas accretion from the circumstellar disk. 
On the other hand, the separation does not increase much if the binary is away from the cloud center (see section~2.2.1 in \citet{Arimoto2021}).

\section{Results}
\label{sec:result}

\subsection{Coalescence Process of a Protostar Binary}

This section presents the results of numerical simulations for the case where the Pop~III protostar binary coalesces. 
We also describe the details of the coalescence process.

\begin{figure}
    \centering
        \includegraphics[width=\linewidth]{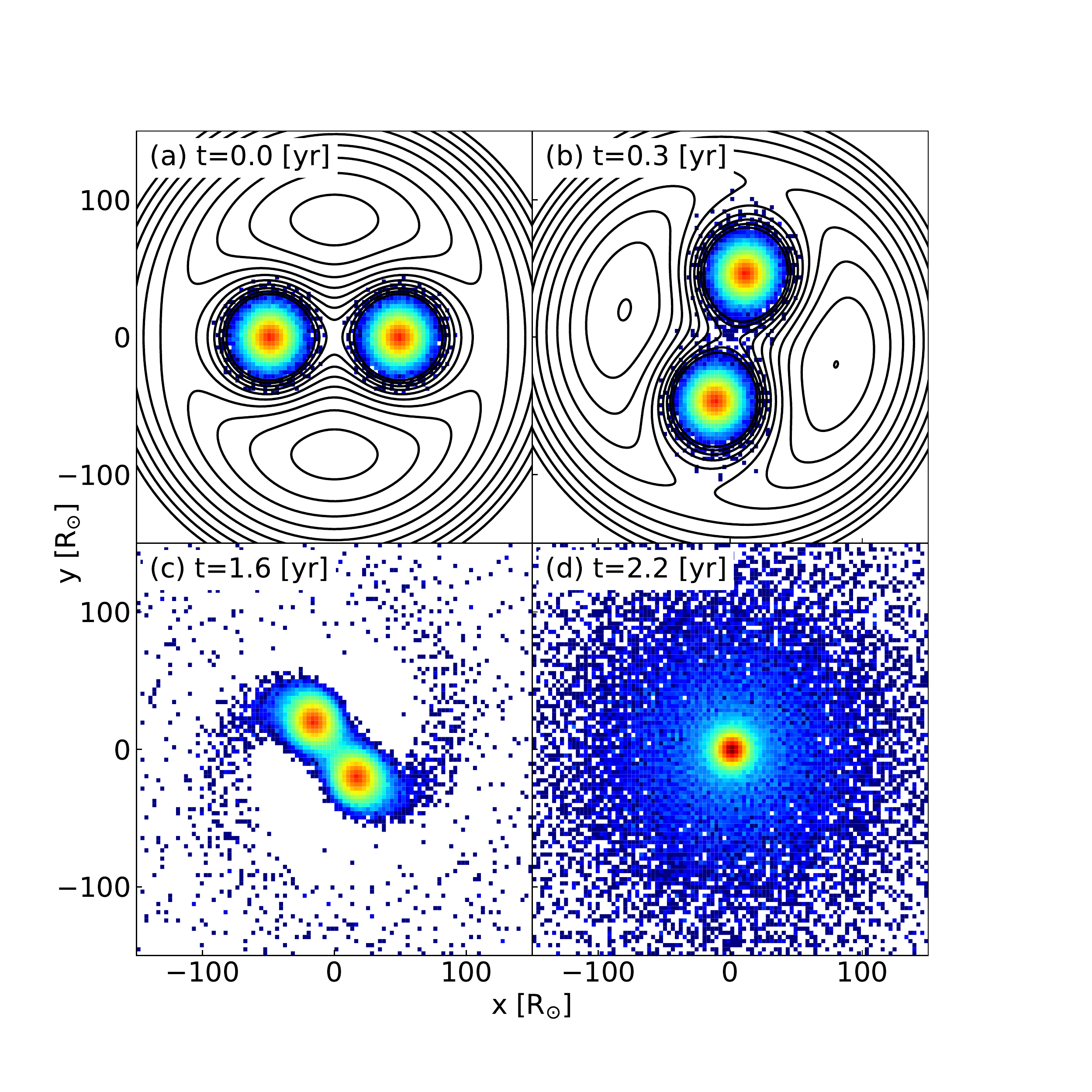}
    \caption{Protostellar gas distribution at (a) $t=0$, (b) 0.3, (c) 1.6, and (d) 2.1~years in model $a_{\rm ini}= 97.9 R_{\odot}$. 
    The black contour lines show the Roche potential of the binary.}
    \label{fig:spatial}
\end{figure}

Figure~\ref{fig:spatial} shows the time evolution of the distribution of the Pop~III protostellar binary in the case of $a_{\rm ini}= 97.9 R_{\odot}$. 
The Roche isopotential surface is overplotted in panels (a) and (b). 
In panel (a), we show the distribution at the beginning of the simulation. Panels (b) and (c) show the snapshots 0.3 and 1.6 years later, respectively. Panel (b) corresponds to the 4th period of the orbital motion, and panel (c) to the 27th period. 
We can see in panel (b) that some particles are about to escape from the outer edge of the binary. 
According to the Roche potential, the region corresponds to the vicinity of the L2 point, where the potential is shallow. 
Then, as the particles escape from the system, the orbital period becomes shorter. 
As shown in panel (c), spiral structures appear just before the coalescence. 
After that, the system immediately coalesces (see panel (d)). 
Panel (d) shows the distribution of the resulting protostar at the time 50~$T_{\rm dyn}$ after the coalescence.

Let us describe the process from panel (a) to (b) in more detail. 
When we start calculating the orbital motion of binary, the protostars begin to spin due to angular momentum transport from the orbital motion by the gravitational torque (detailed analysis is described in Section~\ref{subsec:spinrot}). 
The stars keep spinning up, and their spin angular velocity matches their orbital angular velocity eventually, at which the system is tidally locked. 
This tidal interaction effectively reduces the orbital angular momentum, and thus the separation between the binary stars becomes smaller. 
Gas with a large angular momentum near the L2 points is gradually ejected from the system until the tidal lock is achieved. 
This is because the protostars spin up due to the tidal effect caused by angular momentum transport, and the gas that exceeds the escape velocity can easily escape from the vicinity of the L2 points.

\subsection{Effect of Different Initial Binary Separations}

\begin{figure}
    \centering
        \includegraphics[width=\linewidth]{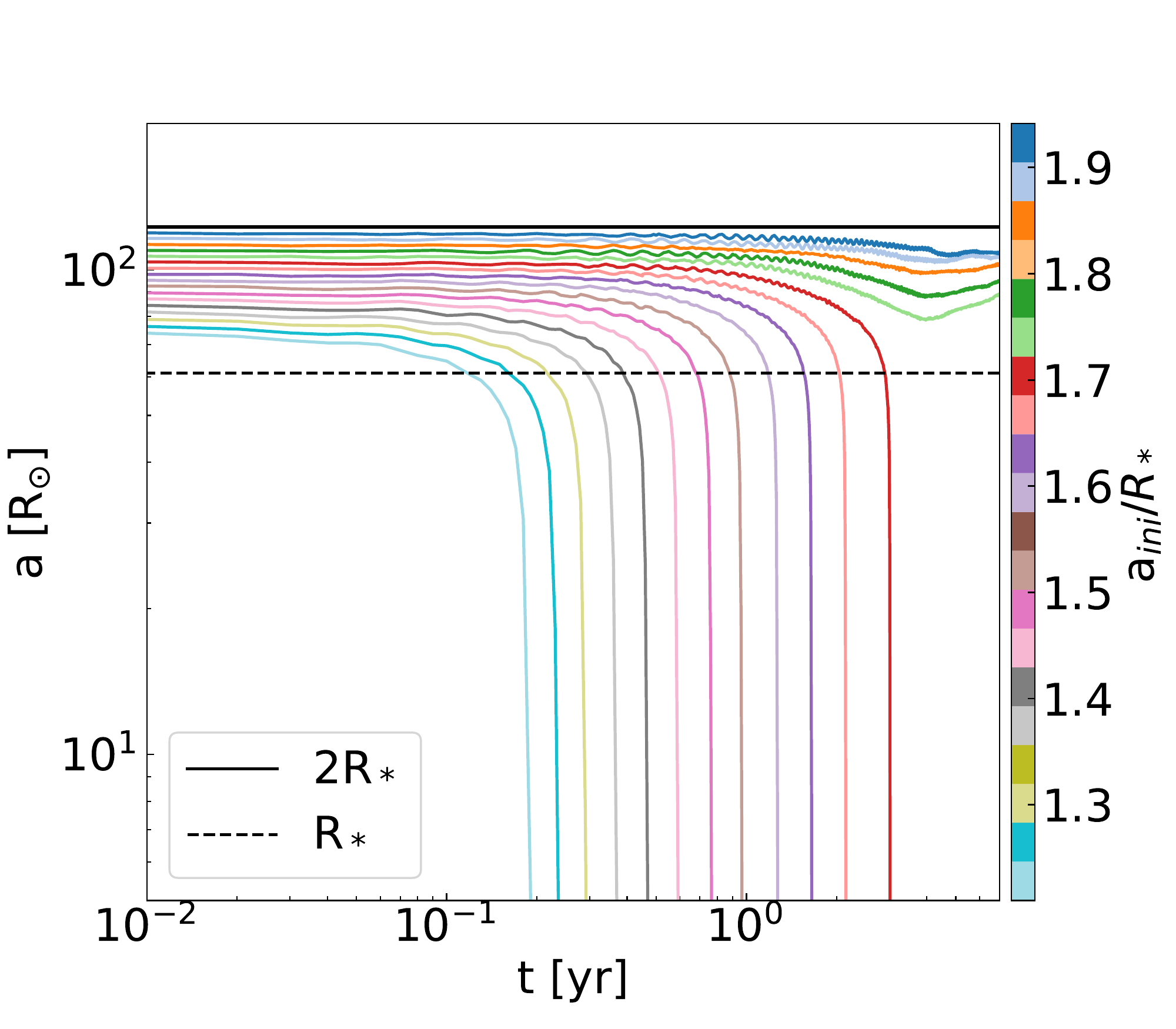}
    \caption{Time evolution of the binary separation. 
    The line colors represent different cases with different initial binary separations. 
    The black dashed and solid lines represent the initial protostellar radius and its doubled separation, respectively.}
    \label{fig:separation}
\end{figure}

Figure~\ref{fig:separation} shows the time evolution of the binary separation $a$. Each line represents the difference in the initial binary separation $a_{\rm ini}$. 
If the initial separation exceeds $106\;R_{\odot}$, the binary does not merge. The time to coalescence varies at approximately equal intervals in logarithmic space for the coalesced parameters. 
In other words, it means that the timescale for coalescence increases exponentially as the initial separation of the binary increases. 
A more detailed analysis of the coalescence time is described in the following section. 

We demonstrate that binaries can coalesce even when the initial binary separation is between the protostar radius and twice its radius. 
As shown in Figure~\ref{fig:IC}, protostars have a two-layer structure, a high-density core and a low-density envelope. 
Our simulations show that the binaries merge when the envelopes overlap to some extent.

\subsection{Timescale of Coalescence}
\label{sec:timescale}

\begin{figure}
    \centering
        \includegraphics[width=\linewidth]{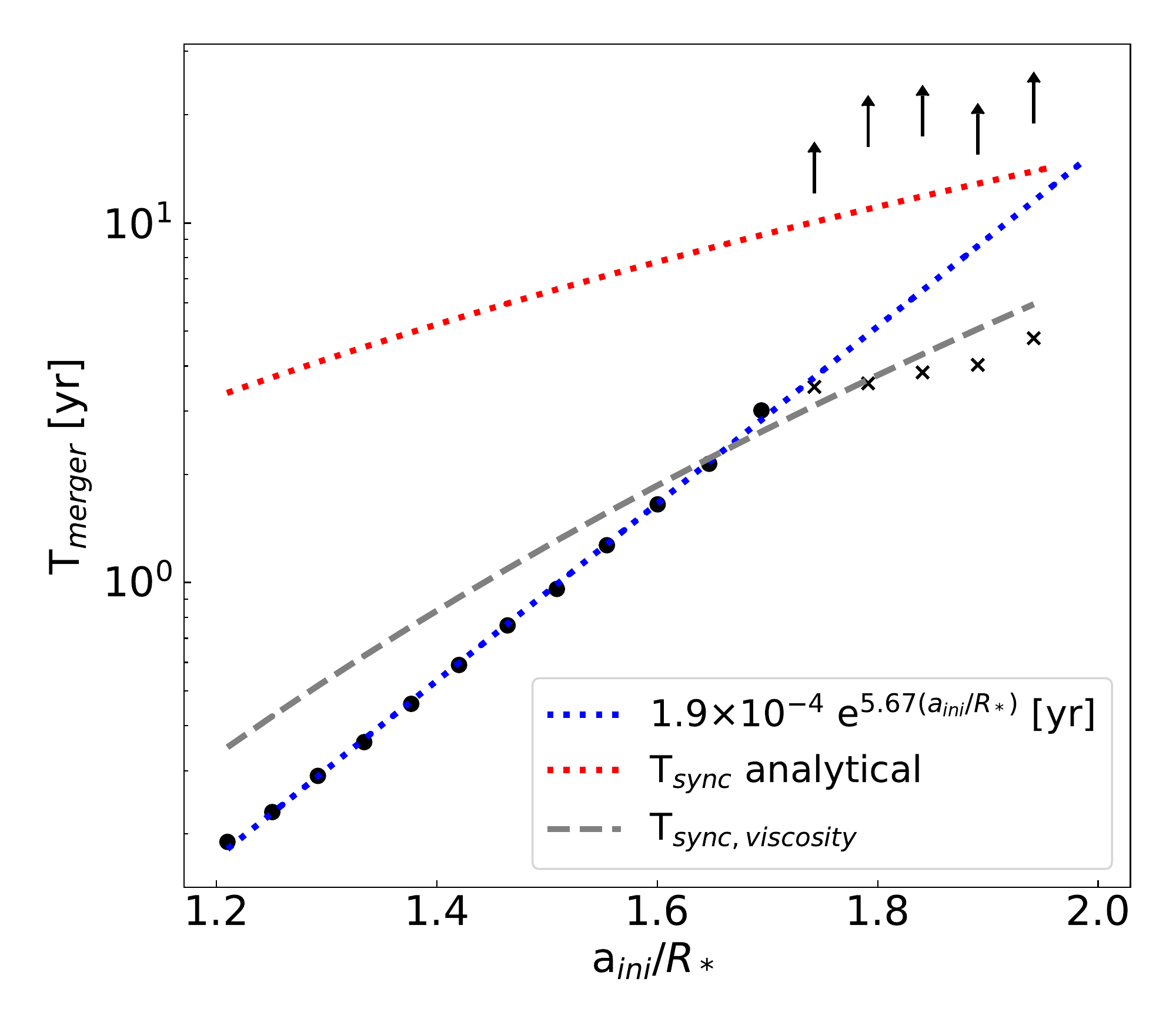}
    \caption{The coalescence time $T_{\rm merger}$ as a function of the initial separation $a_{\rm ini}$ normalized by the protostellar radius. 
    The black points indicate the epochs when the initial binaries coalesce. 
    The black arrows indicate that we compute up to the epoch of the starting point of the arrow without observing stellar coalescence. 
    The black crosses represent the epoch when $\Omega_{\rm spin}/\Omega_{\rm orb}>0.99$ is achieved. 
    The blue dotted line is the least-squares fitting of the black points, where $T_{\rm merger}=1.9\times 10^{-4}\;{\rm e}^{5.67 (a_{\rm ini}/R_*)}$ yr. 
    The red dotted line shows the analytical estimate of the tidal lock timescale. 
    The gray dashed line shows the timescale of momentum transport due to the numerical viscosity. 
}
    \label{fig:fvalue}
\end{figure}

This section discusses the relationship between the time to coalescence and the level of overlap between the protostars. 
Figure~\ref{fig:fvalue} shows the coalescence time $T_{\rm merger}$ as a function of the initial separation $a_{\rm ini}$ normalized by the protostellar radius. 
We define $T_{\rm merger}$ as the epoch when the distance between the center of two protostars approaches less than 30 $R_{\odot}$, which corresponds to half of the original protostellar radius. 
After this, the protostars collapse to a single star immediately, as shown in Figure~\ref{fig:separation}. 
In Figure~\ref{fig:fvalue}, the black points indicate the epochs when the initial binaries coalesce. 
The black arrows indicate that we compute up to the epoch of the starting point of the arrow without observing stellar coalescence. 
The blue dotted line is the least-squares fitting of the black points, where $T_{\rm merger}=1.9\times 10^{-4}\;{\rm e}^{5.67 (a_{\rm ini}/R_*)}$ yr. 
The boundary between merging or not is $a_{\rm ini}/R_*=1.7$, which corresponds to a 15 per~cent overlap between protostars. 
If the initial separation is further apart, the protostars do not merge even after long time calculations. 

\begin{figure}
    \centering
        \includegraphics[width=\linewidth]{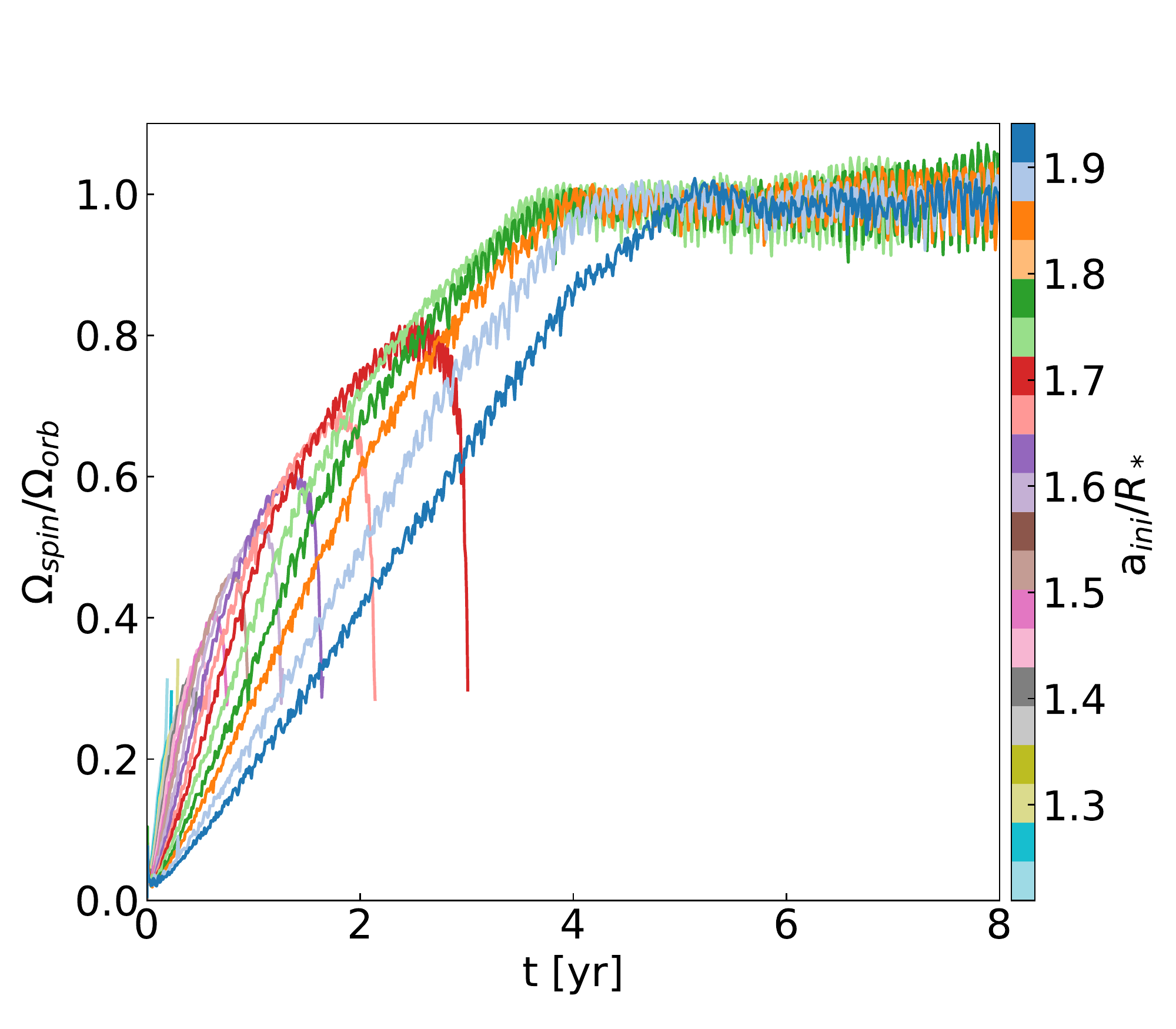}
    \caption{Time evolution of the ratio of protostellar spin angular velocity and orbiting angular velocity as a function of the elapsed time. 
    Each color represents a difference in the initial binary separation.
}
    \label{fig:spinrot}
\end{figure}

For a binary to merge, the orbital angular momentum must be subtracted. 
In the two-body problem, it is known that the orbital angular momentum can be transported to the spin of the protostars by the tidal interaction. 
We show the time evolution of the protostellar spin angular velocity ratio to the orbital angular velocity in Figure~\ref{fig:spinrot}. 
The protostars spin up during their orbital motion and synchronize with the orbital angular velocity. 
We can see that the orbital angular momentum is transported to the stellar spin, thereby reducing the separation of the binary. 
If the initial separation is close enough, the binary merges before the spin-up is completed. 
On the other hand, once $\Omega_{\rm spin}/\Omega_{\rm orb}=1$ is achieved, the orbital angular momentum cannot be transported anymore, and the binary separation cannot be further reduced. 
This mechanism determines the boundary of whether merging will occur or not. 

The black crosses in Figure~\ref{fig:fvalue} represent the epochs when $\Omega_{\rm spin}/\Omega_{\rm orb}>0.99$ is achieved. 
The synchronization timescale of the equal-mass binary can be estimated by 
\begin{equation}
    \frac{1}{T_{\rm sync}}=3\left(\frac{k}{T}\right)\frac{M_*R_*^2}{I}\left(\frac{R_*}{a_{\rm ini}}\right)^6, \label{eq:1_Tsync}
\end{equation}
where $k$, $T$, and $I$ are the apsidal motion constant, the tidal timescale, and the moment of inertia, respectively \citep{Rasio1996,Hurley2002}. 
In the case of the present protostellar model, $M_*R_*^2/I\sim20.3$. 
According to \citet{Hurley2002}, 
\begin{equation}
    \frac{k}{T}=\frac{2}{21}\frac{f_{\rm corr}}{\tau_{\rm c}}\frac{M_{\rm conv}}{M_*}\;{\rm yr^{-1}}, \label{eq:kT}
\end{equation}
where $f_{\rm corr}$ and $M_{\rm conv}$ are the correction of the tidal torque and the mass of the convective layer, respectively. 
The eddy turnover timescale is
\begin{equation}
\tau_{\rm c}= 0.4311\left[\frac{M_{\rm conv}R_{\rm conv}(R_*-\frac{1}{2}R_{\rm conv})}{3L_*}\right]^{1/3},
\end{equation}
where $R_{\rm conv}=58.2\;R_{\odot}$ is the radius of the convective layer, and $L_*$ is the stellar luminosity. 
The protostar consists of a convective structure within a thin radiative layer; therefore, $M_{\rm conv}\approx M_*$. 
The numerical factor $f_{\rm corr}$ is given as 
\begin{equation}
    f_{\rm corr}={\rm min}\left[1,\;\left(\frac{P}{2\tau_c}\right)^2\right],
\end{equation}
where $P$ is the orbital period. 
Only the eddies that turn over in a time less than the tidal timescale ($\sim P$) will contribute to transfer momentum so that both typical length and velocity are reduced by a factor $P/(2\tau_{\rm c})$. 
Now, $\tau_{\rm c}$ is 0.16, and the red dotted line in Figure~\ref{fig:fvalue} shows $T_{\rm sync}$ as a function of the initial separation.

However, the SPH simulations are unable to dissipate the angular momentum with convection within protostars. 
Instead, numerical viscosity is responsible for the angular momentum transport inside the stars. 
Since the relative velocities of particles to $i$-particle within the smoothing length $h_i$ are less than the sound speed $c_i$, we consider only the effect of $\alpha$ of the artificial viscosity. 
The timescale for transporting momentum $\tau_{\rm vis}$ can be expressed as
$\tau_{\rm vis}={R_*}/{v_{\rm vis}}$,
where $v_{\rm vis}$ indicates the velocity of transporting the momentum, which can be written in the following form (see Appendix~\ref{sec:SPHnotation} for the details), 
\begin{equation}
v_{\rm vis}\sim\frac{1}{N}\displaystyle\sum_{i=1}^{N} \frac{1}{4}\alpha c_i\left(\frac{h_i}{r_i}\right)^2. 
\end{equation}
When estimating the synchronization timescale, we replace $\tau_{\rm c}$ in Equation~(\ref{eq:kT}) with $\tau_{\rm vis}$. 
We also set $f_{\rm corr}=1$ because we estimate it by considering the forces that constantly work between particles due to viscosity. 
Therefore the synchronization timescale is
\begin{equation}
    \frac{1}{T_{\rm sync}}=\frac{2}{7}\left(\frac{1}{\tau_{\rm vis}}\right)\frac{M_*R_*^2}{I}\left(\frac{R_*}{a_{\rm ini}}\right)^6. \label{eq:1_Tsync_fin}
\end{equation}

In the case of the SPH method, $\tau_{\rm vis}$ depends on the number of particles; therefore, $T_{\rm sync}$ has a resolution dependence in numerical simulations. 
As the resolution improves, $h$ becomes smaller, and thus viscosity works less effectively. 
If the number of particles representing one protostar is 32768 and 65536, the values of $\tau_{\rm vis}$ are  0.65 and 1.01, respectively. 
Since the smoothing length roughly decreases with $\propto N^{1/3}$, $v_{\rm vis}$ increases with $\propto N^{2/3}$. 
Then, $\tau_{\rm vis}$ becomes $\sim 1.6$ times larger when the number of particles doubles. We further discuss the results by varying the number of particles in Appendix~\ref{sec:convergence}.

If $a_{\rm ini}$ is large, the synchronization timings agree with $T_{\rm sync, viscosity}$. 
As shown in Figure~\ref{fig:spinrot}, the runs with coalesced parameters do not reach $\Omega_{\rm spin}/\Omega_{\rm orb}=1$. 
The binaries have coalesced because they have extracted enough orbital angular momentum from the binary system in a shorter time than it takes to synchronize. 
At the parameter of the coalescing boundary ($a_{\rm ini}/R_*=1.7$), the $T_{\rm sync}$ of the analytical estimate is 9.3 years, which is 3.6 times larger than the $T_{\rm sync, viscosity}$ ($=2.6$ years). 
Therefore, it would require $\sim$220,000 particles to adjust to the timescale of the momentum transport obtained from the analytical estimate of $T_{\rm sync}$. 
Since the numerical cost is extremely high for computation with such a number of particles, we investigate the coalescence mechanism with a lower-resolution simulation in the present study. 
The analytically estimated tidal lock timescale is several times longer than the one obtained by numerical simulation. 
Nevertheless, it is still sufficiently shorter than the KH timescale ($\equiv GM_*^2/R_*L_*\sim 4.1\times10^3$~yr) and other timescales involved in stellar evolution. 
Therefore, the phenomena observed in our simulations leading up to the binary coalescence should not change. 
We confirm that the results for the coalescing parameter ($a_{\rm ini}/R_*<1.7$) converge in the higher-resolution simulations (Appendix~\ref{sec:convergence}).

\subsection{Mass Loss and Angular Momentum Transport}
\label{subsec:spinrot} 

\begin{figure}
    \centering
        \includegraphics[width=\linewidth]{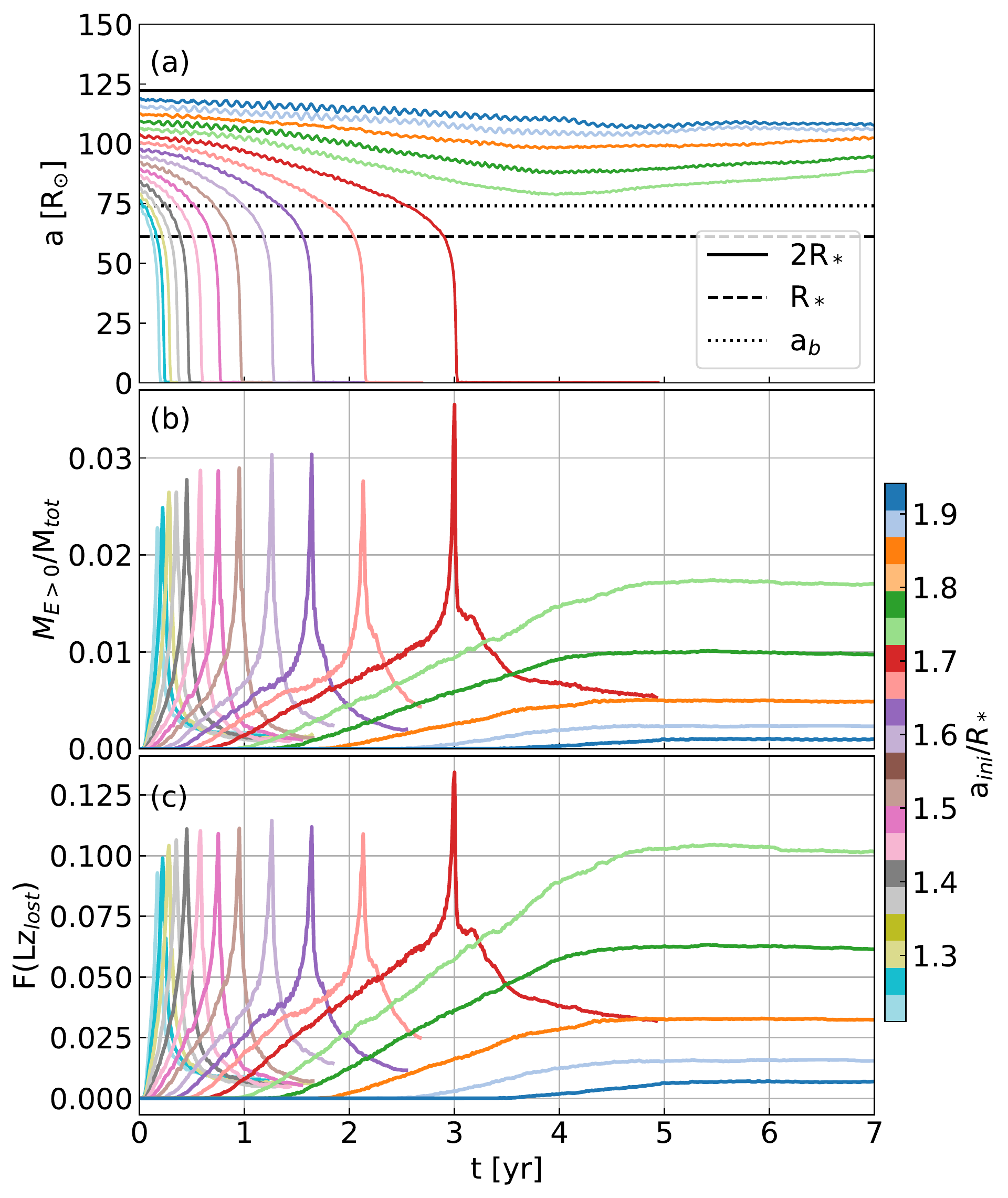}
    \caption{Time evolution of (a) the binary separation, (b) the fraction of unbound mass, and (c) the fraction of angular momentum in the unbound SPH particles. 
    Each color represents a difference in the initial binary separation. 
    The top panel is the same as Figure~\ref{fig:separation}, but both axes are set to a linear scale. 
    The black dotted line in panel (a) shows $a_{\rm b}$. 
    }
    \label{fig:lost}
\end{figure}

Figure~\ref{fig:lost} shows the time evolution of (a) separation, (b) mass loss (fraction of unbound mass), and (c) angular momentum loss (fraction of angular momentum associated with the unbound mass). 
Each color represents the runs with the assumed initial binary separation $a_{\rm ini}$. 
There is a clear difference between the binary that merges, and the one does not. 
For the coalescing ones, when the separation shrinks, the mass of $E_{\rm tot}>0$ increases rapidly (the peaks in panels (b) and (c)). 
Panel (c) shows the time evolution of the ratio of the angular momentum of the $E_{\rm tot}>0$ gas and the total angular momentum of the system. 
A part of gas in the protostar temporarily goes unbound ($E_{\rm tot}>0$) at the moment of coalescence but immediately becomes bounded again through the relaxation process of the gravitational potential. 
Therefore, the mass that is $E_{\rm tot}>0$ after merging decreases. 
If we look at the models that do not coalesce, the smaller the initial separation is, the larger the mass-loss and angular momentum loss is. 
In other words, the binaries that can barely merge are the most likely to suffer mass loss even larger than the merged cases. 
As a result, the gravitational forces between the two stars become weaker. 
Angular momentum is also lost in the no-coalesce cases, but it is insufficient to counteract the reduced gravitational pull. 

To understand the coalesce boundary, we focus on the instantaneous increase in the mass loss just before the coalesce (see Fig.~\ref{fig:lost}(b)). 
At this epoch, the velocity of the tidally spin-up outer edge shakes off the gravitational binding. 
We denote the binary separation and the angular velocity at the boundary as $a_{\rm b}$ and $\Omega_{\rm b}$, respectively. 
Since the velocity of the outer edge of the binary is $(R_*+a_{\rm b}/2)\Omega_{\rm b}$, the balance with the gravitational energy on the mass element of the outer edge is
\begin{equation}
-\frac{GM}{R_*}-\frac{GM}{(a_{\rm b}+R_*)}+\frac{1}{2}\left(R_*+\frac{a_{\rm b}}{2}\right)^2\Omega_{\rm b}^2=0. \label{eq:balance}
\end{equation}
Since $\Omega_{\rm spin}=\Omega_{\rm orb}$ at $a_{\rm b}$, $\Omega_{\rm b}=\sqrt{2GM/{a_{\rm b}^3}}$. 
Note that we ignore the mass loss until the binary separation has reached $a_{\rm b}$. 
Solving the fourth-order equation resulting from Equation~(\ref{eq:balance}), the real solution in the range $a_{\rm b}>0$ is $a_{\rm b} \sim 1.21~R_*$. 
The black dotted line in panel (a) of Figure~\ref{fig:lost} shows $a_{\rm b}$. 
For the no-coalesce cases, the binary separation cannot reach $a_{\rm b}$, and the binary stars recoil. 
In the coalescing parameters, we can see from panel (b) that the mass loss suddenly increases when the binary separation becomes smaller than $\sim a_{\rm b}$. 
The ratio of spin angular velocity to orbiting angular velocity starts to decrease in this phase (see Figure~\ref{fig:spinrot}). 
The angular momentum loss associated with the mass loss leads to a sudden decrease in the binary separation, which causes a rapid increase in $\Omega_{\rm orb}$. 
Since the timescale of this process is sufficiently short compared to the synchronization timescale, the ratio decreases in the phase just before coalescence.

Darwin instability is known for the relationship between stellar spin and orbital angular momentum associated with tidal lock \citep{Hut1980}. 
The condition that prevents binary stars from coalescing due to the stability is 
$a>\sqrt{3(I_1+I_2)/\mu_m}\sim 0.54 R_*$. 
Here, $I_1=I_2=I$, and the reduced mass $\mu_m=M_*/2$. 
However, this is a discussion of whether the condition of tidal lock is valid in a conservative system of both angular momentum and mass. 
In this study, tidal lock causes spin-up of the stars and mass loss. 
Therefore, even if it is stable in terms of Darwin instability, there is no problem with the protostellar coalescence as in the present study.

It is worth noting that we can not resolve the outer part ($> 45~R_{\odot}$) of the protostars well because of the very low gas density (Section~\ref{sec:method}, last paragraph). 
Thus one might suspect the maximal $a_{\rm ini}$ for the merger is underestimated since the unresolved gas in the outer part of the star can take away more angular momentum from the system. 
However, the mass that should reside $45\;R_\odot< r <R_*$, where no SPH particles we have, is only 0.11 per~cent of the stellar mass, as noticed in Section~\ref{sec:method}. 
On the other hand, Figure~\ref{fig:lost}(b) shows that $\sim 1$ per~cent of the total mass is lost along with $\sim$10 per~cent of the angular momentum even in the no-coalescence runs. 
Thus, the additional removal of the angular momentum due to the loss of the very rarefied outer envelope (0.11 per~cent) is unlikely to change the merger condition of the binary. 
To confirm the effect of the numerical resolution, we additionally conduct SPH simulations with a large number of particles as described in Appendix~\ref{sec:convergence}. 

We run numerical simulations with constant mass resolution. 
The equal-mass SPH particles are less effective in modeling the outer layers because we put more particles in the center. 
A constant number density of particles would be better for cases near the maximal $a_{\rm ini}$. 
However, a constant number density may not give the final structure of the coalesced star as accurate. Therefore, each method has its own advantages and disadvantages, and our simulations better represent the stellar structure after coalescence. 

We also investigate the energy conservation of the simulations in the case of low-resolution runs. 
The simulations are performed in an adiabatic system, and the energy error is typically less than 0.5 per~cent until coalescence and 2 per~cent in the worst case. 
Although not used in our simulations, the introduction of "grad-h" corrections \citep[e.g.][]{Springel2010} should improve energy conservation. 
This is because both energy and entropy are integrated exactly when the "grad-h" correction is applied. 
Although there is room for improvement, energy conservation in our simulations is relatively satisfactory. 

\subsection{The Structure of the Coalesced Star}
\label{subsec:CoalecedStar}

\begin{figure}
    \centering
         \includegraphics[width=\linewidth]{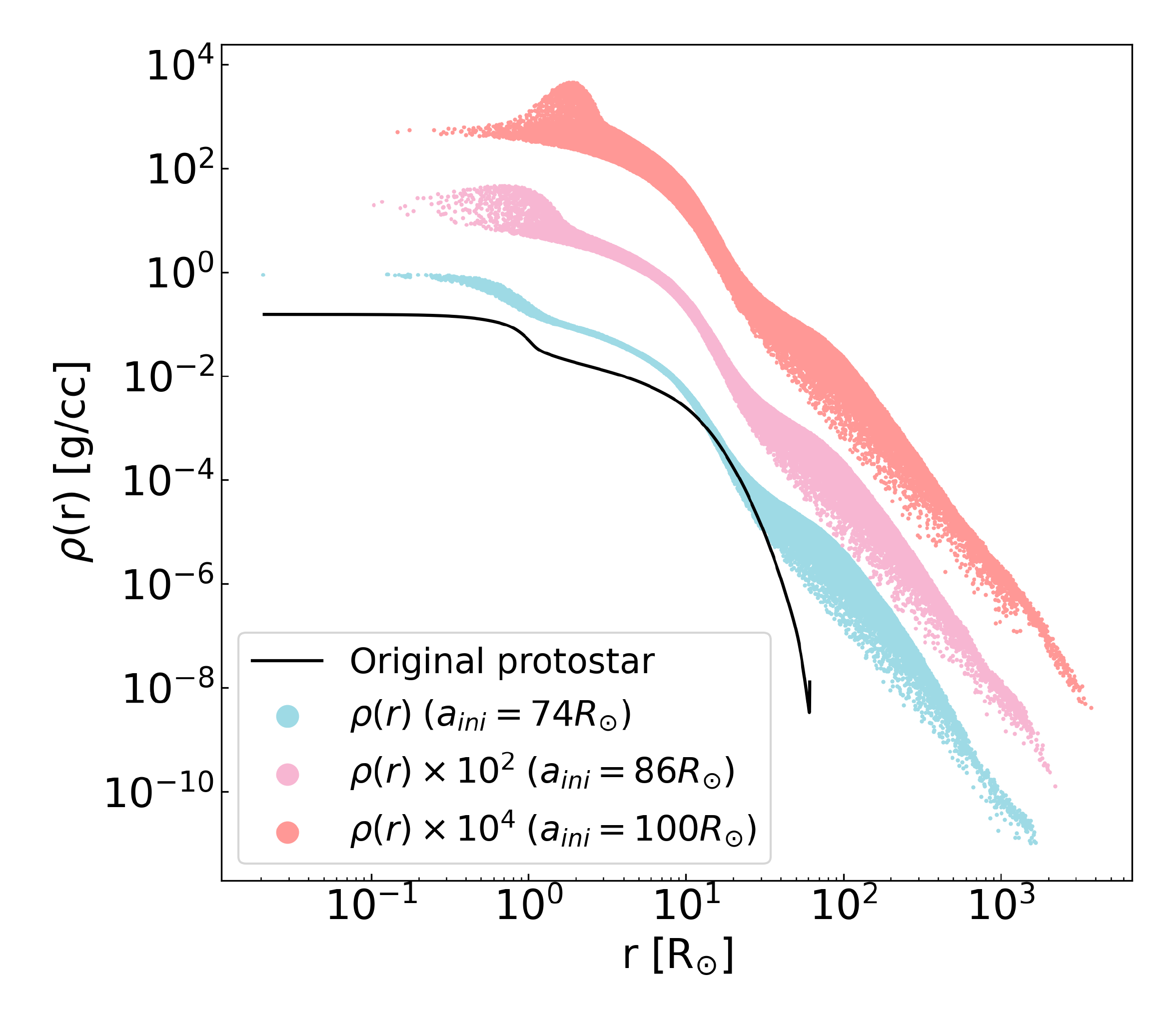}
    \caption{Density profiles at $T_{\rm end}$, with only those particles for which $E_{\rm tot}<0$. 
    The different point colors represent cases with different initial separations $a_{\rm ini}=74\;R_{\odot}$ (cyan), $a_{\rm ini}=86\;R_{\odot}$ (pink), and $a_{\rm ini}=100\;R_{\odot}$ (orange). 
    The pink and orange points are shifted to the larger by two and four orders of magnitude, for clarity. 
    The black line shows the density distribution of the original protostar model. 
    }
    \label{fig:merged_rho}
\end{figure}

We describe here the resulting stellar structure of the coalesced binary. 
Figure~\ref{fig:merged_rho} shows the density profile at 50~$T_{\rm dyn}$ after coalescence, which we call the epoch $T_{\rm end}$. 
Note that this epoch varies between models and is a quantity that is defined only for the coalesced models. 
We plot only the particles for which $E_{\rm tot}<0$. 
The resulting stars consist of a dense core of a few $R_{\odot}$, an envelope of several tens $R_{\odot}$, and an even less dense and huge envelope. 

\begin{figure}
    \centering
        \includegraphics[width=\linewidth]{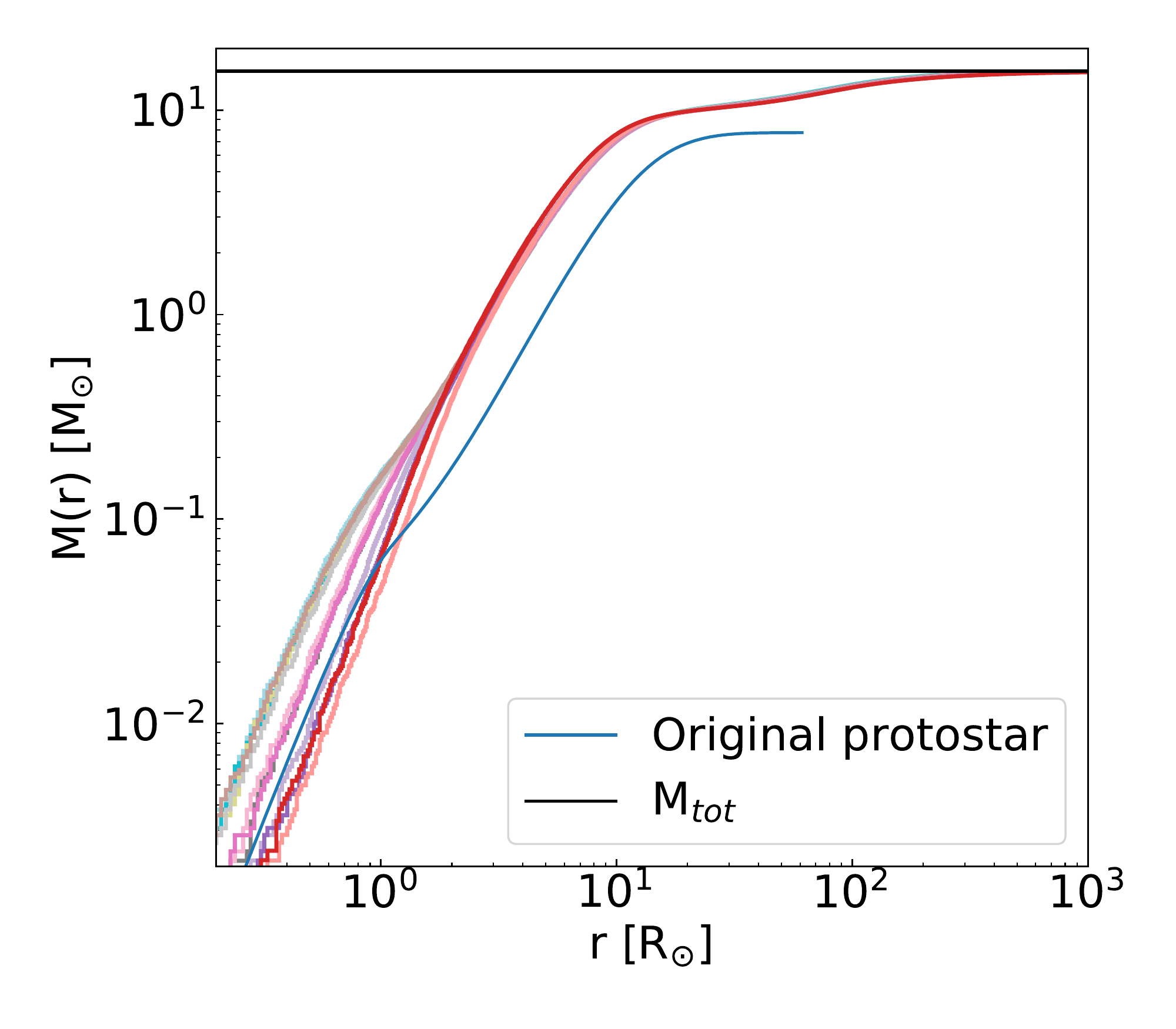}
    \caption{Mass distribution at $T_{\rm end}$, with only those particles for which $E_{\rm tot}<0$. 
    The colors of each line represent the difference in initial separation, which are the same as in Figure~\ref{fig:separation}. 
    The black line indicates the mass twice the original protostellar mass. }
    \label{fig:merged_m}
\end{figure}

Figure~\ref{fig:merged_m} shows the mass distribution at $T_{\rm end}$, with only those particles for which $E_{\rm tot}<0$. 
The mass distributions of the post-merger protostars are all very similar, even though they are slightly different in the inner part of a few $R_{\odot}$. 
The radius of the mass, which exceeds 99 per~cent of the total mass, extends to 200-300~$R_{\odot}$. 
This indicates that the protostellar radius can expand due to the coalescence. 
However, it is hard to follow the evolution for a long time after the coalescence; therefore, it is not clear if the protostellar radius will be maintained for a long time. 
The difficulty is because the increase of the central density drastically shortens the time step of the numerical simulations. 
We also have to consider radiative cooling if we try to follow much further. 
In future work, it is necessary to follow the long-term evolution of stellar structure using radiation hydrodynamics simulation. 

\begin{figure}
    \centering
        \includegraphics[width=\linewidth]{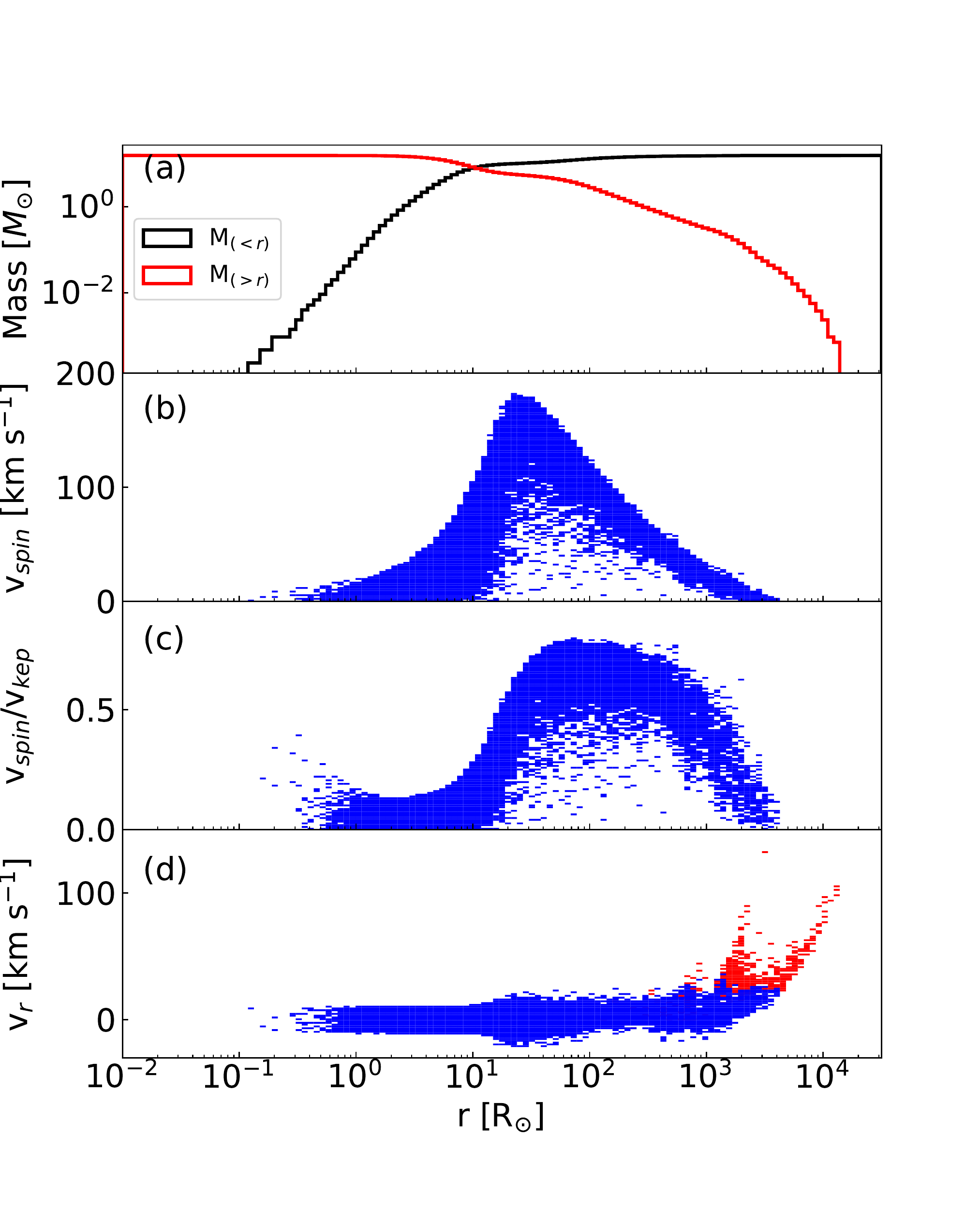}
    \caption{Distributions of (a) cumulative mass, (b) spin velocity, (c) spin velocity normalized by the Keplerian velocity, and (d) radial velocity of the coalesced star. 
    We plot the result at $T_{\rm end}$ for the case of $a_{\rm ini}=103\;R_{\odot}$. 
    The black and red lines show the cumulative and inverse cumulative mass distribution, respectively. 
    The blue region in panels (b)-(d) shows the velocity distribution of the gas bound to the coalesced star. 
    The red region in panel (d) shows the velocity distribution of the gas for $E_{\rm tot}>0$. 
    }
    \label{fig:spin}
\end{figure}

We now describe the rotation of the resulting stars after the coalesce. 
We find that the stars are rapidly rotating in all models at the end of the calculation. 
Panel (b) of Figure~\ref{fig:spin} shows the spin velocity distribution. 
The rotation velocity peaks at $\sim 30$ $R_{\odot}$ (approximately 180 ${\rm km\;s}^{-1}$) and gradually decreases outside the radius. 
We only show the case of $a_{\rm ini}=103\;R_{\odot}$, but the results are almost independent of the initial separation. 
Near the peak, the rotation velocity is around 75 per~cent of the Keplerian velocity (see panel (c)). 
Thus, we find that the post-merger star, which is an outcome of the merger of two initially non-rotating protostars, rotates rapidly.

Panel (d) of Figure~\ref{fig:spin} shows the radial velocity distribution. 
The gas bound and unbound to the star are plotted in blue and red, respectively. 
The unbound gas escapes from the system, accounting for less than five per~cent of the total mass (see also panel (c) in Fig.~\ref{fig:lost}). 
We can see two spike structures in the outflow gas. 
The outer component is ejected gradually during the spin-up of the stars, while the inner component is ejected at the coalescence. 
The outer component has a distribution extended in the direction of the orbital plane of the protostellar binary. 
On the other hand, the inner component has a nearly spherically symmetric distribution. 
It has recently been suggested that the outflows observed in massive star-forming regions may originate from protostellar coalescences \citep{Bally2020,Bally2022}. 
These results imply that future observations will reveal such two-component outflows.

\section{Discussion}
\label{sec:discussion}

\subsection{Implications of Massive Contact Binaries}

In this section, we discuss close binary formation along with the result that tight binaries cannot avoid coalescence during the protostellar phase with large protostellar radii. 
Massive close binaries have been observed in the Galaxy \citep[e.g.][]{Mason2009,Sana2012}. 
In particular, the recently discovered VFTS~352 is a remarkably tight binary with stellar masses of 28.63 $M_{\odot}$ and 28.85 $M_{\odot}$ and with a radius of 7.22 $R_{\odot}$ and 7.25 $R_{\odot}$, respectively, and an orbital semimajor axis is 17.55 $R_{\odot}$. 
This system is expected to be in the contact phase \citep{Almeida2015}. 
Although our study does not explicitly focus on Pop~I stars, it is worth noting that the Pop~I and Pop~III protostars have similar interior structures under rapid mass accretion, and the Pop~I protostars also have larger radii than their main-sequence counterparts \citep[e.g.][]{Hosokawa2009,Hosokawa2010}. 
Therefore, our results suggest that the close binaries of Pop~I protostars also cannot avoid merging before they evolve to main-sequence binaries. 
Alternatively, there would be some mechanism to shrink the orbit after the protostellar phase is over.

In order to explain the observed close binaries, it is necessary to tighten the binary from protostellar binaries with a larger separation. 
Previous works have investigated the mechanism of shrinking binary separation \citep[e.g.][]{Kratter2011,Moe2018,Harada21}. 
If considering a triple system, a third companion perturbs and could tighten the orbit of a binary system called the Kozai-Lidov mechanism \citep[e.g.][]{Kozai1962,Lidov1962,Rose2019}. 
Gas accretion and/or the interaction of a binary with its circumbinary disk could shrink binary separation \citep{Bete2002}. 
On the other hand, accretion of the disk gas, which has a high specific angular momentum, may broaden the binary separation \citep{Chon2019}. 
Thus, the origin of massive close binaries remains unknown. 
As our results demonstrate, there is a minimum binary separation at which a binary can survive, even if the close binary is formed in the protostellar phase. 
The protostellar radius decreases with KH contraction during its evolution into a main-sequence star (see Fig.~\ref{fig:Hosokawa}). 
A large amount of angular momentum must be extracted during this process, which is also happening with a single star. 
It may be worth considering a way to transport angular momentum such that the binary separation could also be shrunk while the protostars are shrinking. 
The three-body interaction can tighten the binary after KH contraction or after main-sequence star formation.

The close binary formation is also crucial for the GW source originating from Pop~III stars. 
The issue of whether binary black holes originating from Pop~III stars can be sources of gravitational waves is an attractive window for understanding the formation of Pop~III stars. 
If Pop~III stars have stellar masses $>50 M_{\odot}$, they evolve to red giants, and such binaries easily coalesce by extracting angular momentum in the common envelope phase, even if they are not close binaries. 
On the other hand, if the stellar mass is $<50 M_{\odot}$, they evolve to blue giants, and the stellar radius does not enlarge so much, indicating that close binaries are more likely to coalesce \citep[e.g.][]{Kinugawa2014,Inayoshi2017,Tanikawa2021}. 
In that case, the closer the binary separation is, the more likely it is to coalesce moderately. 
Therefore, it is helpful to determine how close a protostellar binary can be.

\subsection{Stellar Rotation}
\label{subsec:stellar_rotation}

Stellar rotation is invisible in numerical simulations that use sinks or do not resolve the stellar structure sufficiently. 
In the environment of primordial star formation, stars are expected to experience many protostar mergers \citep[e.g.,][]{Greif2011,Susa2019}. 
According to the present results, the angular momentum transfer from the orbital motion to the spin of the protostars is the key for the protostellar binary coalescence. 
In the initial setup of the simulations of this paper, zero-spin of the protostars is assumed. 
This assumption is simple and suitable for understanding the physical processes of mergers but might not be realistic. 
In the actual situation, the accreting matters bring some angular momentum; thereby, the protostars spin to some extent. 
If the initial spin of the protostars is so rapid, the tidal locking will be achieved earlier than the present results. 
On the other hand, if the protostars are initially rotating rapidly, spin angular momentum can transport to orbital angular momentum. 
As a result, the binary separation could widen and stabilize at a certain distance. 
We can see such a recoil phenomenon in our simulation results for the case of just barely avoiding a merger, where the binary separation after the spin-up increases (see Fig.~\ref{fig:lost}(a)). 
If this inverse process is effective, it still requires an overlooked process after protostellar evolution to form close binary main-sequence binaries. 
Consequently, the merging will be less likely to occur in such an environment, although this should be confirmed in future work. 
In any case, the spin of the protostar is an essential quantity for the merger. 
Since we often introduce sink particles, we should trace the spin of the sinks ($\simeq$ protostars) to implement the sink-sink merger properly.

Present results also imply that the members of the survived tight binaries tend to be fast rotators since their spin should saturate to the level of tidal locking. 
If the separation is of the protostellar radius order, the angular velocity of the protostars approaches the break-up value. 
In addition, if the binary merges to form a single star, it should also be a fast rotator, as shown in Section \ref{subsec:CoalecedStar}. 
One key mechanism that could reduce the spin is magnetic braking. 
The effects of the magnetic field in the primordial environment are still uncertain, but it seems more tangled than the present-day counterpart since the turbulent motion amplifies them  \citep[e.g.,][]{Sur2010,Federrath2011,Higashi2021,Sadanari2021,Hirano2022}. 
Since the magnetic braking effects are less important in the tangled geometry of B-fields, Pop~III stars tend to be fast rotators. 
We also have to quantify the effects of magnetic braking of the tangled B-field in future studies.

When such a resulting fast-rotating massive star evolves and explodes as a supernova, the elemental abundance of the ejecta can change from the case without rotation \citep[e.g.,][]{Choplin2020}. 
Consequently, the metal abundance pattern incorporated into second-generation star formation will be changed. 
For instance, while non-rotating Pop III stars rarely produced nitrogen in supernova explosions, fast rotators could produce it \citep[e.g.,][]{Meynet2002,Takahashi2018}. 
Such spinning Pop~III stars can describe the nitrogen abundance in the galactic extremely metal-poor stars \citep[e.g.,][]{Meynet2006} as well as the abundance of the extremely metal-poor galaxies \citep[e.g.,][]{Goswami2022} nicely. 
In addition, if the supernova explosion energy is sufficiently large, the yield gas can escape the gravitational binding of the parent halos. 
Such metals are ejected into the intergalactic medium \citep{Jaacks2018,Kirihara2020,Liu2020b} and leave some impact on the metal abundance pattern of the QSO absorption line systems \citep{Cooke2011,Welsh2022}. 

\subsection{Effect of Altering the Protostellar Model}

The spin of each protostar aside, we use a specific protostar model as initial conditions. 
The initial conditions differ if we vary the gas accretion rate or the timing of the switch of the accretion flow from spherical to disk accretion. 
The maximum protostellar radius depends on the timing when the accretion from the polar direction becomes negligible \citep{Hosokawa2009, Haemmerle16, Hosokawa2016}. 
Considering the numerical cost, we employ a model with a relatively sizable maximal expansion radius, in which the relatively later switch of the accretion geometry is assumed. 
So far, the timing of the change of accretion flow is uncertain, so it is worthwhile to investigate the binary evolution with a different model (timing) in future studies.

A variety of situations could emerge that differ from this work with respect to the basic parameters of the binaries. 
We can consider the cases of binaries with unequal masses and non-zero eccentricities. 
Thus, we obviously need more studies to sweep the parameter space in the future. 
We note, however, that under massive gas accretion flows, matters generally land more heavily on low mass protostar since they are in outer orbits intersecting the flow earlier. 
As a result, the binaries tend to approach equal mass. 
In fact, observed short-period binaries are typically equal-mass binaries \citep{Lucy1979}. 
We also note that the tidal effect enables them to get close to $e = 0$, even when the orbital eccentricity is non-zero initially \citep{Hurley2002}. 
Most observed short-period binaries have circular orbits, indeed \citep{Duchene2013}.

\section{Summary}
\label{sec:summary}

We have investigated the coalescence condition of Pop~III protostar binaries, which is indispensable in considering the possibility that some evolve into very tight binaries. 
If possible, Pop~III stars are nominated as an origin of gravitational-wave sources. 
Stellar evolution calculations predict that accreting Pop III protostars have radii much larger than their main-sequence counterparts. 
Such a star consists of a high-density core and a vast low-density envelope. 
Therefore, it is essential to investigate how close protostars can approach each other but not coalesce, considering the realistic interior structure.

To investigate the criteria for the coalescence of binaries, we conduct a suite of SPH simulations following the binary evolution. 
We employ a Pop~III protostellar model with the mass and radius of $7.75\;M_{\odot}$ and radius of $61.1\;R_{\odot}$, which is obtained by stellar evolution calculations including effects of the mass accretion \citep{Omukai2003, Hosokawa2010}. 
We establish a protostellar model having the same interior structure in three dimensions. 
We assume an equal-mass binary consisting of these twin stars as the initial condition for the SPH simulations. 
We set the initial orbital eccentricity to be 0. 
We vary the initial separation in the range from $74\;R_{\odot}$ to $118\;R_{\odot}$ at the equal interval of $3\;R_{\odot}$, performing 17 simulation runs in total.

Our simulation results show that the binary separation reduces by the transport of orbital angular momentum to the spin of the protostars by the tidal interaction. 
When the initial separation is smaller than about 80 per~cent of the sum of the protostellar radii, coalescence occurs on a considerably shorter timescale than that of protostellar evolution. 
The mass loss up to the coalescence is less than about three per~cent. 
The remnant of the merger is a rapidly rotating star. 
We find that the structure of the post-merger star is not significantly dependent on the initial separation. Our results suggest that close binaries of protostars cannot avoid merging before evolving to main-sequence binaries.

Our study is also applicable for the massive close binary formation in the present-day universe, where observations confirm the existence of massive binaries with small separations of $< 100\;R_{\odot}$ \citep[e.g.][]{Sana2012,Almeida2015}.
Stellar evolution calculations also predict that Pop~I massive protostars should also inflate to have large radii $\sim 100\;R_{\odot}$, similar to the Pop~III cases, under rapid mass accretion \citep[e.g.][]{Hosokawa2009}. 
This suggests that there are some orbital shrinking mechanisms after the protostars contract to become the zero-age main-sequence stars, which may be operative in both Pop~I and III star formation.


\section{acknowledgments}

We are grateful to the anonymous referee for the useful comments that improved this paper. 
We thank S. Chon, K. Kawaguchi, K. Omukai, K. Tanaka, A. Tanikawa, and D. Toyouchi for fruitful discussions. 
This work was supported by JSPS KAKENHI Grant Number JP17H02869, JP17H06360, JP19H01934, JP21H00041, JP22K03689, and JP22K14076. 
Numerical computations was in part carried out on Cray XC50 at the Center for Computational Astrophysics, National Astronomical Observatory of Japan. 

Numerical analyses were partially carried out using the following packages, NumPy \citep{vanderWalt2011} and matplotlib \citep{Hunter2007}.

\appendix

\section{Timescale of Momentum Transport due to Viscosity}
\label{sec:SPHnotation}

Here, we derive the timescale of momentum transport due to viscosity in the SPH simulations. 
The viscosity term for the equation of motion is 
\begin{equation}\label{eq:motion_app}
  m_i \frac{d^2 \boldsymbol r_i}{dt^2} =-m_i \sum_{j} m_j \Pi_{ij} \nabla_i W_{ij},
\end{equation}
where $W_{ij}$ is a kernel function. 
The artificial viscosity term $\Pi_{ij}$ is
\begin{equation}\label{eq:pi_app}
\Pi_{ij}=
\begin{cases}
  \frac{-\alpha c_{ij} \mu_{ij} +\beta \mu_{ij}^2}{{\rho_{ij}}} & \mbox{if $\boldsymbol{v}_{ij}\cdot\boldsymbol{r}_{ij}<0$} \\
0 & \mbox{otherwise}, \\
\end{cases}
\end{equation}
where 
\begin{equation}
  \mu_{ij}=\frac{{h_{ij}}\,\boldsymbol{v}_{ij}\cdot\boldsymbol{r}_{ij} }
     {\left|\boldsymbol{r}_{ij}\right|^2 + \epsilon h_{ij}^2}.\nonumber
\end{equation}
Here $h_{ij}$, ${\rho_{ij}}$, and ${c_{ij}}$ denote arithmetic means of the smoothing length, density, and sound speed for the two particles $i$ and $j$, respectively, whereas $\boldsymbol{r}_{ij}\equiv \boldsymbol{r}_i - \boldsymbol{r}_j$ and $\boldsymbol{v}_{ij}\equiv \boldsymbol{v}_i - \boldsymbol{v}_j$. 
The strength of the viscosity is controlled by the parameters $\alpha$ and $\beta$. 
The constant $\epsilon$ is introduced to avoid the singularities, and its fiducial value is $\simeq 0.01$.

\begin{figure}
    \centering
    \includegraphics[width=0.5\linewidth]{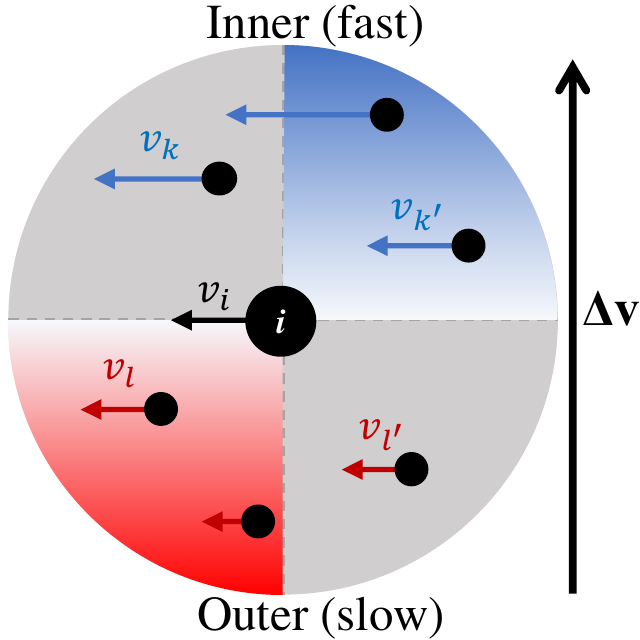}
    \caption{Schematic picture of the motion of the particles within the smoothing length of an $i$-particle. 
    Black filled circles represent the SPH particles. 
    The arrows originating from the circles represent velocity vectors. 
    The upper (lower) side rotates faster (slower) than the central $i$-particle. 
} 
    \label{fig:schemAppendixA}
\end{figure}

Figure~\ref{fig:schemAppendixA} is a schematic picture of the motion of the particles within the smoothing length of an $i$-particle. 
We consider a situation where the host protostar is orbiting, and the upper side is in the direction of the center of the binary system. 
Then, the upper (lower) side rotates faster (slower) than the central $i$-particle. 
In this work, the relative velocities of particles to $i$-particle within the smoothing length $h_i$ are less than the sound speed $c_i$.
Therefore, we only consider the effect of $\alpha$ of the artificial viscosity in the following argument. 
The numerical viscosity only works when the $i$ and $j$ particles approach each other. 
In other words, the artificial viscosity works only between the $k'$-particles and the $i$-particle (blue region in Fig.~\ref{fig:schemAppendixA}) and between the $l$-particles and the $i$-particle (red region). 
The artificial viscosity does not work between the $k$-particles and the $i$-particle and between the $l'$-particles and the $i$-particle. 
Substituting Equation~(\ref{eq:pi_app}) into Equation~(\ref{eq:motion_app}) based on the above situation setting, we have
\begin{eqnarray}
  m_i \frac{d^2 \boldsymbol r_i}{dt^2} & = & -m_i \sum_{j} m_j \frac{-\alpha c_{ij} \mu_{ij}}{{\rho_{ij}}} \nabla_i W_{ij} \nonumber\\
  & = & -m_i \sum_{l} m_l \frac{-\alpha c_{il} \mu_{il}}{{\rho_{il}}} \nabla_i W_{il} 
  +m_i \sum_{k'} m_{k'} \frac{-\alpha c_{ik'} \mu_{ik'}}{{\rho_{ik'}}} \nabla_i W_{ik'}\nonumber\\
  & \simeq & -\frac{m_i}{4} \nabla_i(-\alpha c_{i}\Delta v_{il}) + \frac{m_i}{4} \nabla_i(-\alpha c_{i}\Delta v_{ik'}), \nonumber
\end{eqnarray}
where we suppose $\rho_{ij}\sim\rho_{i}$, $c_{ij}\sim c_{i}$, and $\mu_{ij}\sim\Delta v_{ij}$. 
Since the 1st and the 2nd term on R.H.S. of the above equation almost cancel with each other, each $i$-particle transports the above momentum per unit time.
We consider how many $i$-particles occupy per unit area to obtain the momentum flux $F$. 
Since the separation between particles is $\sim h$, we assume that the unit thickness is $h$. 
We calculate the flux by multiplying this momentum by the particle number density $n$ and the unit thickness $h$. 
\begin{eqnarray}
  F & \simeq & \frac{m_i}{4}\frac{\alpha c_{i}}{r_i}\frac{v_i h_i}{r_i} n_i h_i \nonumber\\
  &=&\frac{\rho_i v_i}{4}\left(\frac{h_i}{r_i}\right)^2 \alpha c_i \nonumber\\
  &\equiv & \rho v_i v_{{\rm vis},i}. \nonumber
\end{eqnarray}
The timescale of momentum transport $\tau_{\rm vis}$ can be expressed as
\begin{eqnarray}
  \tau_{\rm vis} & = & \frac{R_*}{v_{\rm vis}}=\frac{R_*}{\frac{1}{N}\sum_{i} v_{{\rm vis},i}}, \nonumber
\end{eqnarray}
where $v_{\rm vis}$ means the velocity that transfers the momentum. 

\section{Coalescence Timescale with Different Numerical Resolution}
\label{sec:convergence}

\begin{figure}
    \centering
        \includegraphics[width=\linewidth]{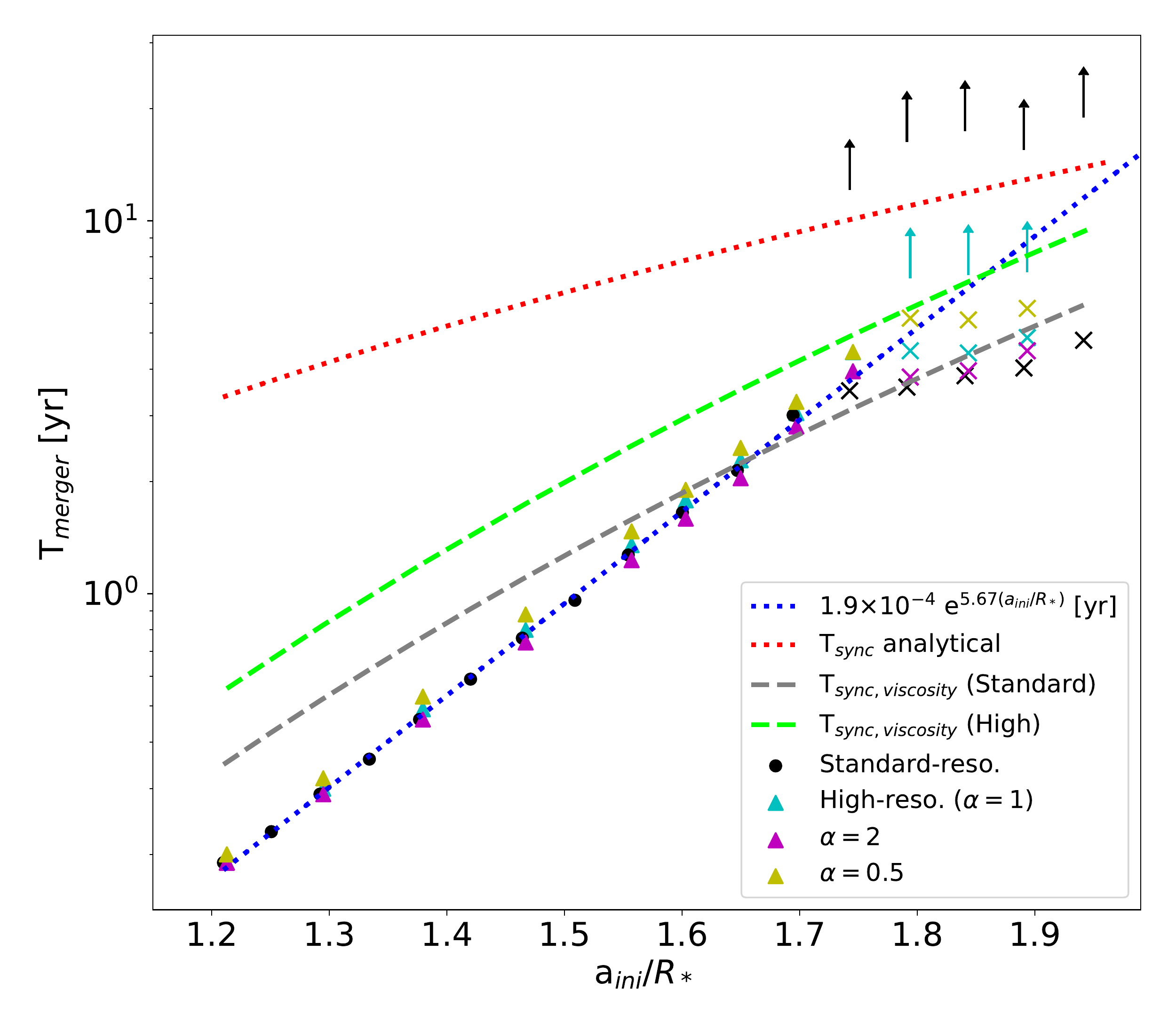}
    \caption{Coalescence time $T_{\rm merger}$ as a function of the initial separation $a_{\rm ini}$ normalized by the protostellar radius with different numerical resolutions. 
    The black circles and cyan triangles indicate the epochs when the binaries coalesce of the standard and high resolution runs, respectively.
    The magenta and yellow symbols correspond to the results of altering the artificial viscosity parameters to $(\alpha, \beta)=(2, 4)$ and $(\alpha, \beta)=(0.5, 1)$. 
    The crosses represent the epochs when $\Omega_{\rm spin}/\Omega_{\rm orb}>0.99$ is achieved. 
    The arrows indicate the cases that do not merge during the computation time. 
    The gray and light green dashed lines show the timescale of momentum transport due to the numerical viscosity in standard- and high- resolution runs, respectively. 
    Other symbols and lines are those of Figure~\ref{fig:fvalue}. 
}
    \label{fig:fvalue_app}
\end{figure}

We perform additional 12 runs of SPH simulations with the number of particles of 65536 (twice the number of the standard resolution runs) to confirm the results with different numerical resolution. 
Figure~\ref{fig:fvalue_app} summarizes the coalescence time as a function of the initial separation for different numerical resolutions. 
The coalescence timescale in higher resolution runs is in good agreement with the standard resolution runs. 
Only at one point near the coalescing boundary, $T_{\rm merger}$ is slightly longer in the case of high-resolution runs.

When we use the SPH method, $\tau_{\rm vis}$ depends on the numerical resolution as described in Section~\ref{sec:timescale}. 
As a result, $T_{\rm sync}$ will change, and the boundary position will likely shift slightly. 
The gray and bright green dashed lines in Figure~\ref{fig:fvalue_app} represent $T_{\rm sync}$ calculated using Equation~(\ref{eq:1_Tsync_fin}) using $\tau_{\rm vis}$ at standard and high resolutions, respectively. 
In fact, $T_{\rm merger}$ is longer and more difficult to merge near the boundary when the number of particles is large. 
This is because $T_{\rm sync}$ becomes longer when the number of particles is larger, making it less likely for viscosity to work. 
At the parameter of the coalescing boundary ($a_{\rm ini}/R_*=1.7$), the $T_{\rm sync}$ of the analytical estimate is 9.3 years, which is 3.6 times larger than the $T_{\rm sync}$ ($=2.6$ years) obtained using the SPH simulation with standard resolution. 
It requires $\sim$220,000 particles to adjust to the timescale of the momentum transport obtained from the analytical estimate of $T_{\rm sync}$. 

To examine the uncertainty of the coalescence boundary due to differences in numerical viscosity, we also run the simulations with the parameter pairs $(\alpha, \beta)=(2, 4)$ and $(\alpha, \beta)=(0.5, 1)$. 
For both parameters, we perform 12 runs of high-resolution SPH simulations. 
We show the coalescence and synchronization time of the additional runs in Figure~\ref{fig:fvalue_app} with the circle and cross symbols. 
As the viscosity parameter increases (decreases), the efficiency of angular momentum transport increases (decreases), leading to a decrease (increase) in the time required for coalescence. 
However, the coalescing boundary remains unaltered by changes in the viscosity parameter. 
This result can be attributed to the constancy of the total amount of angular momentum that needs to be transported before coalescence occurs.



\begin{thebibliography}{}
\expandafter\ifx\csname natexlab\endcsname\relax\def\natexlab#1{#1}\fi
\providecommand{\url}[1]{\href{#1}{#1}}
\providecommand{\dodoi}[1]{doi:~\href{http://doi.org/#1}{\nolinkurl{#1}}}
\providecommand{\doeprint}[1]{\href{http://ascl.net/#1}{\nolinkurl{http://ascl.net/#1}}}
\providecommand{\doarXiv}[1]{\href{https://arxiv.org/abs/#1}{\nolinkurl{https://arxiv.org/abs/#1}}}

\bibitem[{{Abel} {et~al.}(2002){Abel}, {Bryan}, \& {Norman}}]{Abel2002}
{Abel}, T., {Bryan}, G.~L., \& {Norman}, M.~L. 2002, Science, 295, 93,
  \dodoi{10.1126/science.295.5552.93}

\bibitem[{{Almeida} {et~al.}(2015){Almeida}, {Sana}, {de Mink}, {Tramper},
  {Soszy{\'n}ski}, {Langer}, {Barb{\'a}}, {Cantiello}, {Damineli}, {de Koter},
  {Garcia}, {Gr{\"a}fener}, {Herrero}, {Howarth}, {Ma{\'\i}z Apell{\'a}niz},
  {Norman}, {Ram{\'\i}rez-Agudelo}, \& {Vink}}]{Almeida2015}
{Almeida}, L.~A., {Sana}, H., {de Mink}, S.~E., {et~al.} 2015, \apj, 812, 102,
  \dodoi{10.1088/0004-637X/812/2/102}

\bibitem[{{Arimoto} {et~al.}(2021){Arimoto}, {Asada}, {Cherry}, {Fujii},
  {Fukazawa}, {Harada}, {Hayama}, {Hosokawa}, {Ioka}, {Itoh}, {Kanda},
  {Kawabata}, {Kawaguchi}, {Kawai}, {Kobayashi}, {Kohri}, {Koshio}, {Kotake},
  {Kumamoto}, {Machida}, {Matsufuru}, {Mihara}, {Mori}, {Morokuma},
  {Mukohyama}, {Nakano}, {Narikawa}, {Negoro}, {Nishizawa}, {Ohgami}, {Omukai},
  {Sakamoto}, {Sako}, {Sasada}, {Sekiguchi}, {Serino}, {Soda}, {Sugita},
  {Sumiyoshi}, {Susa}, {Suyama}, {Takahashi}, {Takahashi}, {Takiwaki},
  {Tanaka}, {Tanaka}, {Tanikawa}, {Tominaga}, {Uchikata}, {Utsumi}, {Vagins},
  {Yamada}, \& {Yoshida}}]{Arimoto2021}
{Arimoto}, M., {Asada}, H., {Cherry}, M.~L., {et~al.} 2021, arXiv e-prints,
  arXiv:2104.02445.
\newblock \doarXiv{2104.02445}

\bibitem[{{Bally} {et~al.}(2022){Bally}, {Chia}, {Ginsburg}, {Reipurth},
  {Tanaka}, {Zinnecker}, \& {Faulhaber}}]{Bally2022}
{Bally}, J., {Chia}, Z., {Ginsburg}, A., {et~al.} 2022, \apj, 924, 50,
  \dodoi{10.3847/1538-4357/ac30de}

\bibitem[{{Bally} {et~al.}(2020){Bally}, {Ginsburg}, {Forbrich}, \&
  {Vargas-Gonz{\'a}lez}}]{Bally2020}
{Bally}, J., {Ginsburg}, A., {Forbrich}, J., \& {Vargas-Gonz{\'a}lez}, J. 2020,
  \apj, 889, 178, \dodoi{10.3847/1538-4357/ab65f2}

\bibitem[{{Bate} {et~al.}(2002){Bate}, {Bonnell}, \& {Bromm}}]{Bete2002}
{Bate}, M.~R., {Bonnell}, I.~A., \& {Bromm}, V. 2002, \mnras, 336, 705,
  \dodoi{10.1046/j.1365-8711.2002.05775.x}

\bibitem[{{Belczynski} {et~al.}(2017){Belczynski}, {Ryu}, {Perna}, {Berti},
  {Tanaka}, \& {Bulik}}]{Belczynski2017}
{Belczynski}, K., {Ryu}, T., {Perna}, R., {et~al.} 2017, \mnras, 471, 4702,
  \dodoi{10.1093/mnras/stx1759}

\bibitem[{{Benz} \& {Hills}(1987)}]{Benz1987}
{Benz}, W., \& {Hills}, J.~G. 1987, \apj, 323, 614, \dodoi{10.1086/165857}

\bibitem[{{Benz} \& {Hills}(1992)}]{Benz1992}
---. 1992, \apj, 389, 546, \dodoi{10.1086/171230}

\bibitem[{{Chiaki} \& {Yoshida}(2022)}]{Chiaki2022}
{Chiaki}, G., \& {Yoshida}, N. 2022, \mnras, 510, 5199,
  \dodoi{10.1093/mnras/stab2799}

\bibitem[{{Chon} \& {Hosokawa}(2019)}]{Chon2019}
{Chon}, S., \& {Hosokawa}, T. 2019, \mnras, 488, 2658,
  \dodoi{10.1093/mnras/stz1824}

\bibitem[{{Choplin} {et~al.}(2020){Choplin}, {Tominaga}, \&
  {Meyer}}]{Choplin2020}
{Choplin}, A., {Tominaga}, N., \& {Meyer}, B.~S. 2020, \aap, 639, A126,
  \dodoi{10.1051/0004-6361/202037966}

\bibitem[{{Clark} {et~al.}(2011){Clark}, {Glover}, {Smith}, {Greif}, {Klessen},
  \& {Bromm}}]{Clark2011}
{Clark}, P.~C., {Glover}, S.~C.~O., {Smith}, R.~J., {et~al.} 2011, Science,
  331, 1040, \dodoi{10.1126/science.1198027}

\bibitem[{{Cooke} {et~al.}(2011){Cooke}, {Pettini}, {Steidel}, {Rudie}, \&
  {Nissen}}]{Cooke2011}
{Cooke}, R., {Pettini}, M., {Steidel}, C.~C., {Rudie}, G.~C., \& {Nissen},
  P.~E. 2011, \mnras, 417, 1534, \dodoi{10.1111/j.1365-2966.2011.19365.x}

\bibitem[{{Costa} {et~al.}(2023){Costa}, {Mapelli}, {Iorio}, {Santoliquido},
  {Escobar}, {Klessen}, \& {Bressan}}]{Costa2023}
{Costa}, G., {Mapelli}, M., {Iorio}, G., {et~al.} 2023, arXiv e-prints,
  arXiv:2303.15511, \dodoi{10.48550/arXiv.2303.15511}

\bibitem[{{Davies} {et~al.}(1994){Davies}, {Benz}, {Piran}, \&
  {Thielemann}}]{Davies1994}
{Davies}, M.~B., {Benz}, W., {Piran}, T., \& {Thielemann}, F.~K. 1994, \apj,
  431, 742, \dodoi{10.1086/174525}

\bibitem[{{Duch{\^e}ne} \& {Kraus}(2013)}]{Duchene2013}
{Duch{\^e}ne}, G., \& {Kraus}, A. 2013, \araa, 51, 269,
  \dodoi{10.1146/annurev-astro-081710-102602}

\bibitem[{{Federrath} {et~al.}(2011){Federrath}, {Sur}, {Schleicher},
  {Banerjee}, \& {Klessen}}]{Federrath2011}
{Federrath}, C., {Sur}, S., {Schleicher}, D. R.~G., {Banerjee}, R., \&
  {Klessen}, R.~S. 2011, \apj, 731, 62, \dodoi{10.1088/0004-637X/731/1/62}

\bibitem[{{Freitag} \& {Benz}(2005)}]{Freitag2005}
{Freitag}, M., \& {Benz}, W. 2005, \mnras, 358, 1133,
  \dodoi{10.1111/j.1365-2966.2005.08770.x}

\bibitem[{{Goswami} {et~al.}(2022){Goswami}, {Silva}, {Bressan}, {Grisoni},
  {Costa}, {Marigo}, {Granato}, {Lapi}, \& {Spera}}]{Goswami2022}
{Goswami}, S., {Silva}, L., {Bressan}, A., {et~al.} 2022, \aap, 663, A1,
  \dodoi{10.1051/0004-6361/202142031}

\bibitem[{{Greif} {et~al.}(2011){Greif}, {Springel}, {White}, {Glover},
  {Clark}, {Smith}, {Klessen}, \& {Bromm}}]{Greif2011}
{Greif}, T.~H., {Springel}, V., {White}, S.~D.~M., {et~al.} 2011, \apj, 737,
  75, \dodoi{10.1088/0004-637X/737/2/75}

\bibitem[{{Haemmerl{\'e}} {et~al.}(2016){Haemmerl{\'e}}, {Eggenberger},
  {Meynet}, {Maeder}, \& {Charbonnel}}]{Haemmerle16}
{Haemmerl{\'e}}, L., {Eggenberger}, P., {Meynet}, G., {Maeder}, A., \&
  {Charbonnel}, C. 2016, \aap, 585, A65, \dodoi{10.1051/0004-6361/201527202}

\bibitem[{{Haiman} {et~al.}(1996){Haiman}, {Thoul}, \& {Loeb}}]{Haiman1996}
{Haiman}, Z., {Thoul}, A.~A., \& {Loeb}, A. 1996, \apj, 464, 523,
  \dodoi{10.1086/177343}

\bibitem[{{Harada} {et~al.}(2021){Harada}, {Hirano}, {Machida}, \&
  {Hosokawa}}]{Harada21}
{Harada}, N., {Hirano}, S., {Machida}, M.~N., \& {Hosokawa}, T. 2021, \mnras,
  508, 3730, \dodoi{10.1093/mnras/stab2780}

\bibitem[{{Hartwig} {et~al.}(2016){Hartwig}, {Volonteri}, {Bromm}, {Klessen},
  {Barausse}, {Magg}, \& {Stacy}}]{Hartwig2016}
{Hartwig}, T., {Volonteri}, M., {Bromm}, V., {et~al.} 2016, \mnras, 460, L74,
  \dodoi{10.1093/mnrasl/slw074}

\bibitem[{{Hernquist} \& {Katz}(1989)}]{Hernquist1989}
{Hernquist}, L., \& {Katz}, N. 1989, \apjs, 70, 419, \dodoi{10.1086/191344}

\bibitem[{{Higashi} {et~al.}(2021){Higashi}, {Susa}, \& {Chiaki}}]{Higashi2021}
{Higashi}, S., {Susa}, H., \& {Chiaki}, G. 2021, \apj, 915, 107,
  \dodoi{10.3847/1538-4357/ac01c7}

\bibitem[{{Hirano} \& {Machida}(2022)}]{Hirano2022}
{Hirano}, S., \& {Machida}, M.~N. 2022, \apjl, 935, L16,
  \dodoi{10.3847/2041-8213/ac85e0}

\bibitem[{{Hosokawa} {et~al.}(2016){Hosokawa}, {Hirano}, {Kuiper}, {Yorke},
  {Omukai}, \& {Yoshida}}]{Hosokawa2016}
{Hosokawa}, T., {Hirano}, S., {Kuiper}, R., {et~al.} 2016, \apj, 824, 119,
  \dodoi{10.3847/0004-637X/824/2/119}

\bibitem[{{Hosokawa} \& {Omukai}(2009)}]{Hosokawa2009}
{Hosokawa}, T., \& {Omukai}, K. 2009, \apj, 691, 823,
  \dodoi{10.1088/0004-637X/691/1/823}

\bibitem[{{Hosokawa} {et~al.}(2010){Hosokawa}, {Yorke}, \&
  {Omukai}}]{Hosokawa2010}
{Hosokawa}, T., {Yorke}, H.~W., \& {Omukai}, K. 2010, \apj, 721, 478,
  \dodoi{10.1088/0004-637X/721/1/478}

\bibitem[{Hunter(2007)}]{Hunter2007}
Hunter, J.~D. 2007, Computing in Science \& Engineering, 9, 90,
  \dodoi{10.1109/MCSE.2007.55}

\bibitem[{{Hurley} {et~al.}(2002){Hurley}, {Tout}, \& {Pols}}]{Hurley2002}
{Hurley}, J.~R., {Tout}, C.~A., \& {Pols}, O.~R. 2002, \mnras, 329, 897,
  \dodoi{10.1046/j.1365-8711.2002.05038.x}

\bibitem[{{Hut}(1980)}]{Hut1980}
{Hut}, P. 1980, \aap, 92, 167

\bibitem[{{Inayoshi} {et~al.}(2017){Inayoshi}, {Hirai}, {Kinugawa}, \&
  {Hotokezaka}}]{Inayoshi2017}
{Inayoshi}, K., {Hirai}, R., {Kinugawa}, T., \& {Hotokezaka}, K. 2017, \mnras,
  468, 5020, \dodoi{10.1093/mnras/stx757}

\bibitem[{{Jaacks} {et~al.}(2018){Jaacks}, {Thompson}, {Finkelstein}, \&
  {Bromm}}]{Jaacks2018}
{Jaacks}, J., {Thompson}, R., {Finkelstein}, S.~L., \& {Bromm}, V. 2018,
  \mnras, 475, 4396, \dodoi{10.1093/mnras/sty062}

\bibitem[{{Kinugawa} {et~al.}(2014){Kinugawa}, {Inayoshi}, {Hotokezaka},
  {Nakauchi}, \& {Nakamura}}]{Kinugawa2014}
{Kinugawa}, T., {Inayoshi}, K., {Hotokezaka}, K., {Nakauchi}, D., \&
  {Nakamura}, T. 2014, \mnras, 442, 2963, \dodoi{10.1093/mnras/stu1022}

\bibitem[{{Kinugawa} {et~al.}(2016){Kinugawa}, {Miyamoto}, {Kanda}, \&
  {Nakamura}}]{Kinugawa2016}
{Kinugawa}, T., {Miyamoto}, A., {Kanda}, N., \& {Nakamura}, T. 2016, \mnras,
  456, 1093, \dodoi{10.1093/mnras/stv2624}

\bibitem[{{Kinugawa} {et~al.}(2020){Kinugawa}, {Nakamura}, \&
  {Nakano}}]{Kinugawa2020}
{Kinugawa}, T., {Nakamura}, T., \& {Nakano}, H. 2020, \mnras, 498, 3946,
  \dodoi{10.1093/mnras/staa2511}

\bibitem[{{Kirihara} {et~al.}(2020){Kirihara}, {Hasegawa}, {Umemura}, {Mori},
  \& {Ishiyama}}]{Kirihara2020}
{Kirihara}, T., {Hasegawa}, K., {Umemura}, M., {Mori}, M., \& {Ishiyama}, T.
  2020, \mnras, 491, 4387, \dodoi{10.1093/mnras/stz3376}

\bibitem[{{Kozai}(1962)}]{Kozai1962}
{Kozai}, Y. 1962, \aj, 67, 591, \dodoi{10.1086/108790}

\bibitem[{{Kratter}(2011)}]{Kratter2011}
{Kratter}, K.~M. 2011, in Astronomical Society of the Pacific Conference
  Series, Vol. 447, Evolution of Compact Binaries, ed. L.~{Schmidtobreick},
  M.~R. {Schreiber}, \& C.~{Tappert}, 47.
\newblock \doarXiv{1109.3740}

\bibitem[{{Krumholz} \& {Thompson}(2007)}]{Krumholz2007}
{Krumholz}, M.~R., \& {Thompson}, T.~A. 2007, \apj, 661, 1034,
  \dodoi{10.1086/515566}

\bibitem[{{Lidov}(1962)}]{Lidov1962}
{Lidov}, M.~L. 1962, \planss, 9, 719, \dodoi{10.1016/0032-0633(62)90129-0}

\bibitem[{{Liu} \& {Bromm}(2020{\natexlab{a}})}]{Liu2020a}
{Liu}, B., \& {Bromm}, V. 2020{\natexlab{a}}, \mnras, 495, 2475,
  \dodoi{10.1093/mnras/staa1362}

\bibitem[{{Liu} \& {Bromm}(2020{\natexlab{b}})}]{Liu2020b}
---. 2020{\natexlab{b}}, \mnras, 497, 2839, \dodoi{10.1093/mnras/staa2143}

\bibitem[{{Lucy} \& {Ricco}(1979)}]{Lucy1979}
{Lucy}, L.~B., \& {Ricco}, E. 1979, \aj, 84, 401, \dodoi{10.1086/112434}

\bibitem[{{Mason} {et~al.}(2009){Mason}, {Hartkopf}, {Gies}, {Henry}, \&
  {Helsel}}]{Mason2009}
{Mason}, B.~D., {Hartkopf}, W.~I., {Gies}, D.~R., {Henry}, T.~J., \& {Helsel},
  J.~W. 2009, \aj, 137, 3358, \dodoi{10.1088/0004-6256/137/2/3358}

\bibitem[{{McKee} \& {Tan}(2003)}]{McKee03}
{McKee}, C.~F., \& {Tan}, J.~C. 2003, \apj, 585, 850, \dodoi{10.1086/346149}

\bibitem[{{Meynet} {et~al.}(2006){Meynet}, {Ekstr{\"o}m}, \&
  {Maeder}}]{Meynet2006}
{Meynet}, G., {Ekstr{\"o}m}, S., \& {Maeder}, A. 2006, \aap, 447, 623,
  \dodoi{10.1051/0004-6361:20053070}

\bibitem[{{Meynet} \& {Maeder}(2002)}]{Meynet2002}
{Meynet}, G., \& {Maeder}, A. 2002, \aap, 390, 561,
  \dodoi{10.1051/0004-6361:20020755}

\bibitem[{{Moe} \& {Kratter}(2018)}]{Moe2018}
{Moe}, M., \& {Kratter}, K.~M. 2018, \apj, 854, 44,
  \dodoi{10.3847/1538-4357/aaa6d2}

\bibitem[{{Monaghan} \& {Varnas}(1988)}]{Monaghan1988}
{Monaghan}, J.~J., \& {Varnas}, S.~R. 1988, \mnras, 231, 515,
  \dodoi{10.1093/mnras/231.3.515}

\bibitem[{{Nishi} \& {Susa}(1999)}]{Nishi1999}
{Nishi}, R., \& {Susa}, H. 1999, \apjl, 523, L103, \dodoi{10.1086/312277}

\bibitem[{{Omukai} \& {Palla}(2003)}]{Omukai2003}
{Omukai}, K., \& {Palla}, F. 2003, \apj, 589, 677, \dodoi{10.1086/374810}

\bibitem[{{Palla} \& {Stahler}(1992)}]{Palla1992}
{Palla}, F., \& {Stahler}, S.~W. 1992, \apj, 392, 667, \dodoi{10.1086/171468}

\bibitem[{{Prole} {et~al.}(2022){Prole}, {Clark}, {Klessen}, \&
  {Glover}}]{Prole2022}
{Prole}, L.~R., {Clark}, P.~C., {Klessen}, R.~S., \& {Glover}, S. C.~O. 2022,
  \mnras, 510, 4019, \dodoi{10.1093/mnras/stab3697}

\bibitem[{{Rasio} {et~al.}(1996){Rasio}, {Tout}, {Lubow}, \&
  {Livio}}]{Rasio1996}
{Rasio}, F.~A., {Tout}, C.~A., {Lubow}, S.~H., \& {Livio}, M. 1996, \apj, 470,
  1187, \dodoi{10.1086/177941}

\bibitem[{{Rose} {et~al.}(2019){Rose}, {Naoz}, \& {Geller}}]{Rose2019}
{Rose}, S.~C., {Naoz}, S., \& {Geller}, A.~M. 2019, \mnras, 488, 2480,
  \dodoi{10.1093/mnras/stz1846}

\bibitem[{{Sadanari} {et~al.}(2021){Sadanari}, {Omukai}, {Sugimura},
  {Matsumoto}, \& {Tomida}}]{Sadanari2021}
{Sadanari}, K.~E., {Omukai}, K., {Sugimura}, K., {Matsumoto}, T., \& {Tomida},
  K. 2021, \mnras, 505, 4197, \dodoi{10.1093/mnras/stab1330}

\bibitem[{{Sana} {et~al.}(2012){Sana}, {de Mink}, {de Koter}, {Langer},
  {Evans}, {Gieles}, {Gosset}, {Izzard}, {Le Bouquin}, \&
  {Schneider}}]{Sana2012}
{Sana}, H., {de Mink}, S.~E., {de Koter}, A., {et~al.} 2012, Science, 337, 444,
  \dodoi{10.1126/science.1223344}

\bibitem[{{Santoliquido} {et~al.}(2023){Santoliquido}, {Mapelli}, {Iorio},
  {Costa}, {Glover}, {Hartwig}, {Klessen}, \& {Merli}}]{Santoliquido2023}
{Santoliquido}, F., {Mapelli}, M., {Iorio}, G., {et~al.} 2023, arXiv e-prints,
  arXiv:2303.15515, \dodoi{10.48550/arXiv.2303.15515}

\bibitem[{{Sato} {et~al.}(2015){Sato}, {Nakasato}, {Tanikawa}, {Nomoto},
  {Maeda}, \& {Hachisu}}]{Sato2015}
{Sato}, Y., {Nakasato}, N., {Tanikawa}, A., {et~al.} 2015, \apj, 807, 105,
  \dodoi{10.1088/0004-637X/807/1/105}

\bibitem[{{Seidl} \& {Cameron}(1972)}]{Seidl1972}
{Seidl}, F.~G.~P., \& {Cameron}, A.~G.~W. 1972, \apss, 15, 44,
  \dodoi{10.1007/BF00649946}

\bibitem[{{Sharda} {et~al.}(2020){Sharda}, {Federrath}, \&
  {Krumholz}}]{Sharda2020}
{Sharda}, P., {Federrath}, C., \& {Krumholz}, M.~R. 2020, \mnras, 497, 336,
  \dodoi{10.1093/mnras/staa1926}

\bibitem[{{Shima} \& {Hosokawa}(2021)}]{Shima21}
{Shima}, K., \& {Hosokawa}, T. 2021, \mnras, 508, 4767,
  \dodoi{10.1093/mnras/stab2844}

\bibitem[{{Smith} {et~al.}(2011){Smith}, {Glover}, {Clark}, {Greif}, \&
  {Klessen}}]{Smith2011}
{Smith}, R.~J., {Glover}, S. C.~O., {Clark}, P.~C., {Greif}, T., \& {Klessen},
  R.~S. 2011, \mnras, 414, 3633, \dodoi{10.1111/j.1365-2966.2011.18659.x}

\bibitem[{{Springel}(2010)}]{Springel2010}
{Springel}, V. 2010, \mnras, 401, 791, \dodoi{10.1111/j.1365-2966.2009.15715.x}

\bibitem[{{Stacy} {et~al.}(2010){Stacy}, {Greif}, \& {Bromm}}]{Stacy2010}
{Stacy}, A., {Greif}, T.~H., \& {Bromm}, V. 2010, \mnras, 403, 45,
  \dodoi{10.1111/j.1365-2966.2009.16113.x}

\bibitem[{{Stacy} {et~al.}(2012){Stacy}, {Greif}, \& {Bromm}}]{Stacy2012}
---. 2012, \mnras, 422, 290, \dodoi{10.1111/j.1365-2966.2012.20605.x}

\bibitem[{{Stahler} {et~al.}(1986){Stahler}, {Palla}, \&
  {Salpeter}}]{Stahler86}
{Stahler}, S.~W., {Palla}, F., \& {Salpeter}, E.~E. 1986, \apj, 302, 590,
  \dodoi{10.1086/164018}

\bibitem[{{Sugimura} {et~al.}(2020){Sugimura}, {Matsumoto}, {Hosokawa},
  {Hirano}, \& {Omukai}}]{Sugimura2020}
{Sugimura}, K., {Matsumoto}, T., {Hosokawa}, T., {Hirano}, S., \& {Omukai}, K.
  2020, \apjl, 892, L14, \dodoi{10.3847/2041-8213/ab7d37}

\bibitem[{{Sur} {et~al.}(2010){Sur}, {Schleicher}, {Banerjee}, {Federrath}, \&
  {Klessen}}]{Sur2010}
{Sur}, S., {Schleicher}, D.~R.~G., {Banerjee}, R., {Federrath}, C., \&
  {Klessen}, R.~S. 2010, \apjl, 721, L134, \dodoi{10.1088/2041-8205/721/2/L134}

\bibitem[{{Susa}(2006)}]{Susa2006}
{Susa}, H. 2006, \pasj, 58, 445, \dodoi{10.1093/pasj/58.2.445}

\bibitem[{{Susa}(2013)}]{Susa2013}
---. 2013, \apj, 773, 185, \dodoi{10.1088/0004-637X/773/2/185}

\bibitem[{{Susa}(2019)}]{Susa2019}
---. 2019, \apj, 877, 99, \dodoi{10.3847/1538-4357/ab1b6f}

\bibitem[{{Susa} {et~al.}(2014){Susa}, {Hasegawa}, \& {Tominaga}}]{Susa2014}
{Susa}, H., {Hasegawa}, K., \& {Tominaga}, N. 2014, \apj, 792, 32,
  \dodoi{10.1088/0004-637X/792/1/32}

\bibitem[{{Susa} \& {Umemura}(2004)}]{Susa2004}
{Susa}, H., \& {Umemura}, M. 2004, \apj, 600, 1, \dodoi{10.1086/379784}

\bibitem[{{Suzuki} {et~al.}(2007){Suzuki}, {Nakasato}, {Baumgardt},
  {Ibukiyama}, {Makino}, \& {Ebisuzaki}}]{Suzuki2007}
{Suzuki}, T.~K., {Nakasato}, N., {Baumgardt}, H., {et~al.} 2007, \apj, 668,
  435, \dodoi{10.1086/521214}

\bibitem[{{Takahashi} {et~al.}(2018){Takahashi}, {Yoshida}, \&
  {Umeda}}]{Takahashi2018}
{Takahashi}, K., {Yoshida}, T., \& {Umeda}, H. 2018, \apj, 857, 111,
  \dodoi{10.3847/1538-4357/aab95f}

\bibitem[{{Tanikawa} {et~al.}(2021{\natexlab{a}}){Tanikawa}, {Kinugawa},
  {Yoshida}, {Hijikawa}, \& {Umeda}}]{Tanikawa2021}
{Tanikawa}, A., {Kinugawa}, T., {Yoshida}, T., {Hijikawa}, K., \& {Umeda}, H.
  2021{\natexlab{a}}, \mnras, 505, 2170, \dodoi{10.1093/mnras/stab1421}

\bibitem[{{Tanikawa} {et~al.}(2015){Tanikawa}, {Nakasato}, {Sato}, {Nomoto},
  {Maeda}, \& {Hachisu}}]{Tanikawa2015}
{Tanikawa}, A., {Nakasato}, N., {Sato}, Y., {et~al.} 2015, \apj, 807, 40,
  \dodoi{10.1088/0004-637X/807/1/40}

\bibitem[{{Tanikawa} {et~al.}(2021{\natexlab{b}}){Tanikawa}, {Susa}, {Yoshida},
  {Trani}, \& {Kinugawa}}]{Tanikawa2021a}
{Tanikawa}, A., {Susa}, H., {Yoshida}, T., {Trani}, A.~A., \& {Kinugawa}, T.
  2021{\natexlab{b}}, \apj, 910, 30, \dodoi{10.3847/1538-4357/abe40d}

\bibitem[{{Tanikawa} {et~al.}(2022){Tanikawa}, {Yoshida}, {Kinugawa}, {Trani},
  {Hosokawa}, {Susa}, \& {Omukai}}]{Tanikawa2022}
{Tanikawa}, A., {Yoshida}, T., {Kinugawa}, T., {et~al.} 2022, \apj, 926, 83,
  \dodoi{10.3847/1538-4357/ac4247}

\bibitem[{{Tegmark} {et~al.}(1997){Tegmark}, {Silk}, {Rees}, {Blanchard},
  {Abel}, \& {Palla}}]{Tegmark1997}
{Tegmark}, M., {Silk}, J., {Rees}, M.~J., {et~al.} 1997, \apj, 474, 1,
  \dodoi{10.1086/303434}

\bibitem[{{Tokovinin} \& {Moe}(2020)}]{Tokovinin2020}
{Tokovinin}, A., \& {Moe}, M. 2020, \mnras, 491, 5158,
  \dodoi{10.1093/mnras/stz3299}

\bibitem[{{van der Walt} {et~al.}(2011){van der Walt}, {Colbert}, \&
  {Varoquaux}}]{vanderWalt2011}
{van der Walt}, S., {Colbert}, S.~C., \& {Varoquaux}, G. 2011, Computing in
  Science and Engineering, 13, 22, \dodoi{10.1109/MCSE.2011.37}

\bibitem[{{Welsh} {et~al.}(2022){Welsh}, {Cooke}, {Fumagalli}, \&
  {Pettini}}]{Welsh2022}
{Welsh}, L., {Cooke}, R., {Fumagalli}, M., \& {Pettini}, M. 2022, \apj, 929,
  158, \dodoi{10.3847/1538-4357/ac4503}

\bibitem[{{Wollenberg} {et~al.}(2020){Wollenberg}, {Glover}, {Clark}, \&
  {Klessen}}]{Wollenberg2020}
{Wollenberg}, K. M.~J., {Glover}, S. C.~O., {Clark}, P.~C., \& {Klessen}, R.~S.
  2020, \mnras, 494, 1871, \dodoi{10.1093/mnras/staa289}

\bibitem[{{Yorke} \& {Bodenheimer}(2008)}]{Yorke2008}
{Yorke}, H.~W., \& {Bodenheimer}, P. 2008, in Astronomical Society of the
  Pacific Conference Series, Vol. 387, Massive Star Formation: Observations
  Confront Theory, ed. H.~{Beuther}, H.~{Linz}, \& T.~{Henning}, 189

\bibitem[{{Yoshida} {et~al.}(2003){Yoshida}, {Abel}, {Hernquist}, \&
  {Sugiyama}}]{Yoshida2003}
{Yoshida}, N., {Abel}, T., {Hernquist}, L., \& {Sugiyama}, N. 2003, \apj, 592,
  645, \dodoi{10.1086/375810}

\end{thebibliography}



\end{document}